\documentclass[
reprint,
twocolumn,
aps,
amsmath,
amssymb,
superscriptaddress,
notitlepage,
longbibliography,
nofootinbib
]{revtex4-1}

\usepackage{lipsum}
\usepackage{hyperref}
\usepackage{graphicx}
\usepackage{amsmath, amsfonts, amssymb}
\usepackage{braket}
\usepackage{dsfont}
\usepackage{upgreek}
\usepackage{qcircuit}
\usepackage{tikz}
\usetikzlibrary{matrix}

\usepackage[utf8]{inputenc}
\usepackage[english]{babel}
\usepackage{hyperref}
\usepackage{array}
\usepackage{mathtools}
\usepackage{booktabs}

\newcommand{\transpose}{\intercal}
\newcommand{\id}{\mathds{1}}
\newcommand{\vac}{\ket{\emptyset}}
\newcommand{\ketbra}[2]{
    \lvert #1 \rangle \! \langle #2 \rvert
}
\newcommand{\adag}{\hat{a}^{\dagger}}
\newcommand{\bdag}{\hat{b}^{\dagger}}

\newcommand{\cz}{\text{c}\sigma_{z}}
\newcommand{\czbar}{\overline{\text{c}\sigma_{z}}}
\newcommand{\cx}{\text{c}\sigma_{x}}

\newcommand{\smat}[4]{
    \scriptsize
    \begin{pmatrix}
    #1 & #2 \\
    #3 & #4
    \end{pmatrix}
}

\newcommand{\tab}[1]{\begin{tabular}{c} #1 \end{tabular}}

\DeclarePairedDelimiter{\abs}{\lvert}{\rvert}
\DeclarePairedDelimiter{\ceil}{\lceil}{\rceil}
\DeclarePairedDelimiter{\floor}{\lfloor}{\rfloor}

\newcolumntype{C}{>{$}c<{$}}
\AtBeginDocument{
    \heavyrulewidth=.08em
    \lightrulewidth=.05em
    \cmidrulewidth=.03em
    \belowrulesep=.65ex
    \belowbottomsep=0pt
    \aboverulesep=.4ex
    \abovetopsep=0pt
    \cmidrulesep=\doublerulesep
    \cmidrulekern=.5em
    \defaultaddspace=.5em
}

\begin{document}

\title{Universal programmable photonic architecture for quantum information processing}

\author{Ben Bartlett}
\email{benbartlett@stanford.edu}
\affiliation{Department of Applied Physics, Stanford University, Stanford, CA 94305, USA}

\author{Shanhui Fan}
\email{shanhui@stanford.edu}
\affiliation{Department of Electrical Engineering, Stanford University, Stanford, CA 94305, USA}

\begin{abstract}
We present a photonic integrated circuit architecture for a quantum programmable gate array (QPGA) capable of preparing arbitrary quantum states and operators. The architecture consists of a lattice of phase-modulated Mach-Zehnder interferometers, which perform rotations on path-encoded photonic qubits, and embedded quantum emitters, which use a two-photon scattering process to implement a deterministic controlled-$\sigma_z$ operation between adjacent qubits. By appropriately setting phase shifts within the lattice, the device can be programmed to implement any quantum circuit without hardware modifications. We provide algorithms for exactly preparing arbitrary quantum states and operators on the device and we show that gradient-based optimization can train a simulated QPGA to automatically implement highly compact approximations to important quantum circuits with near-unity fidelity.
\end{abstract}

\maketitle

\section{Introduction}

There has been growing interest in universal photonic devices which can be dynamically reconfigured to implement any linear optical transformation to a set of coherent optical modes. \cite{Reck1994ExperimentalOperator, Miller2013Self-configuringComponent, Clements2016AnInterferometers, Harris2018LinearProcessors} These devices are often implemented as a mesh of phase-modulated Mach-Zehnder interferometers (MZIs) which can be configured progressively \cite{Reck1994ExperimentalOperator} or simultaneously \cite{Pai2019ParallelNetwork} to apply arbitrary unitary transformations to an input vector of spatial modes. Such devices have a wide range of applications in classical information processing \cite{Harris2018LinearProcessors, Miller2015SortingLight, Annoni2017UnscramblingModes, Zhuang2015ProgrammableApplications, Perez2017MultipurposeCore, Perez2018ProgrammableNanophotonics}, and integrated universal photonic circuits provides an especially promising hardware platform for high-throughput, energy-efficient machine learning. \cite{Shen2017DeepCircuits, Hughes2018TrainingBackpropagation, Pai2019MatrixDevices, Williamson2020ReprogrammableNetworks} 

These devices also have promising applications in quantum information processing: recent demonstrations of boson sampling \cite{Spring2013BosonChip}, quantum transport dynamics \cite{Harris2017QuantumProcessor}, photonic quantum walks \cite{Grafe2016IntegratedWalks}, counterfactual communication \cite{AlonsoCalafell2019Trace-freeProcessor}, and probabilistic two-photon gates \cite{Carolan2015UniversalOptics} have all been performed on this type of programmable photonic hardware. Photonic systems offer a range of unique advantages over other substrates for quantum information processing: optical quantum states have long coherence times and can be maintained at room temperature, since they interact very weakly with their environment; photonic qubits are optimal information carriers for distant nodes within quantum networks; and MZIs provide simple, high-fidelity implementations of single-qubit operations which can be integrated into a photonic chip.

However, photonic quantum computation poses several intrinsic difficulties. The non-interacting nature of photons makes implementing deterministic multi-photon quantum gates a challenge; many existing proposals \cite{Knill2001AOptics} and demonstrations \cite{Carolan2015UniversalOptics} of linear optical quantum computing rely on non-deterministic ``heralded'' gates, or encode multi-qubit quantum states in exponentially many spatial modes \cite{Russell2017DirectMatrices}. Since photons must propagate at the speed of light, photonic quantum processing must be done along the path of the photon by sequential optical components, making complex quantum circuits prohibitively large to implement with free-space optics. These systems and even some integrated photonic circuits also often suffer from a lack of reconfigurability, as the design of task-specific optical circuity must be modified to perform different computations. \cite{Takeda2019TowardComputing}

Here we describe a photonic lattice architecture for a reconfigurable and universal quantum programmable gate array (QPGA) which can implement any quantum operation, in principle deterministically and with perfect fidelity. Our design, described in Section \ref{sec:lattice_design}, is similar to a universal linear optical component, but employs nonlinear interactions from precisely placed quantum emitters to enable an $N$-qubit state to be encoded using $\mathcal{O}(N)$ number of spatial modes. The proposed device can be programmed to implement any quantum circuit decomposed as one- and two-qubit gates which could physically implemented by lattice components on an integrated photonic circuit. Phase-modulated MZIs apply arbitrary single-qubit operations to qubits which are path-encoded by single photons in a superposition of pairs of waveguides, two-photon scattering processes induced by strongly-coupled quantum emitters implement controlled gates between adjacent qubits. 

We provide exact algorithms in Section \ref{sec:implementing_quantum_operations} for obtaining the appropriate phase shifter parameters to prepare arbitrary quantum states and operators on-chip. In Section \ref{sec:gradient-based-circuit-optimization}, we discuss how optimization techniques from machine learning can be used to automatically discover high-fidelity approximations to desired quantum operations which are significantly more compact than their explicitly-decomposed exact representations.

\begin{figure*}[ht]
    \centering
    \includegraphics[width=\textwidth]{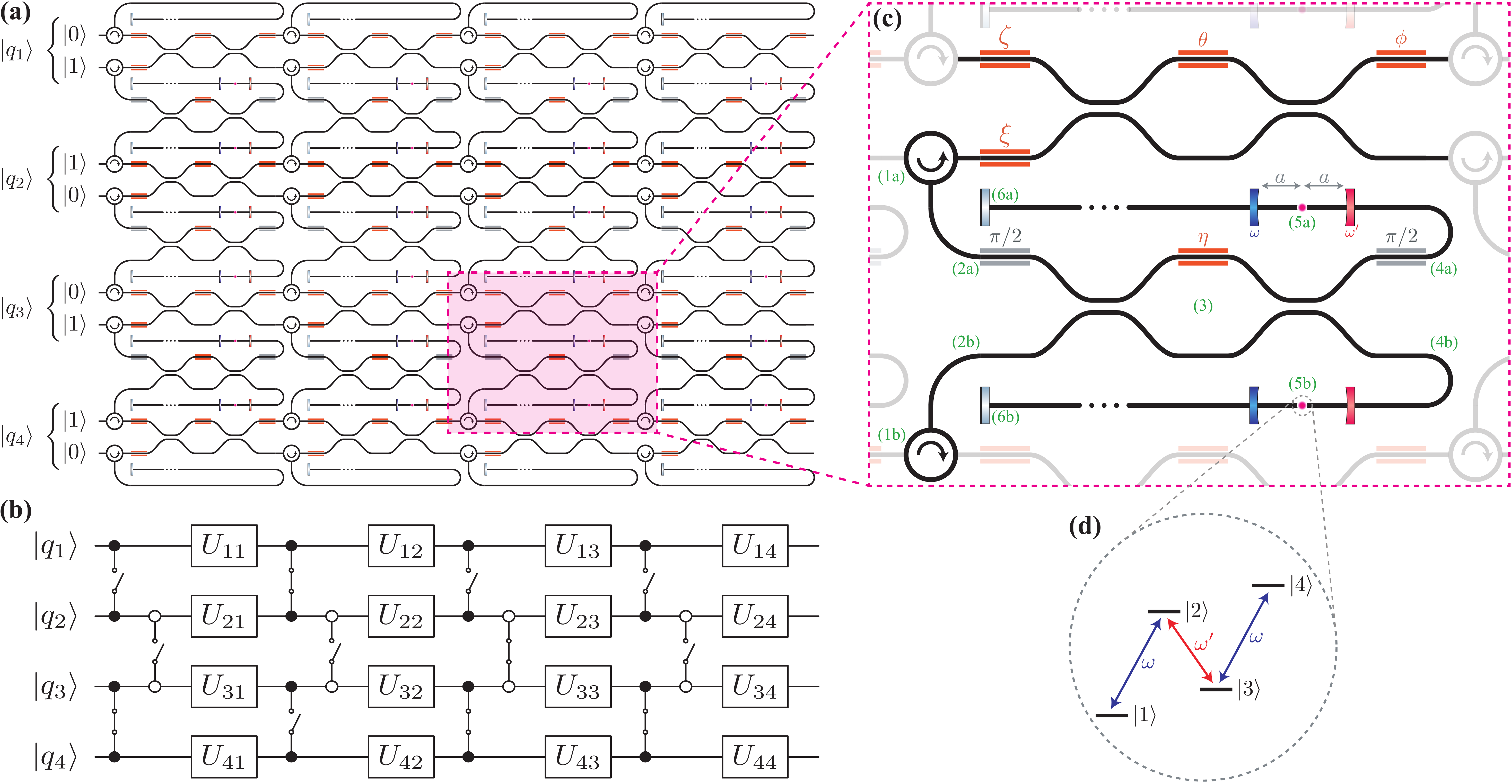}
    \caption{The architecture for the quantum programmable gate array shown at various levels of detail. 
    (a) Physical layout of a four-qubit QPGA with a depth of four layers. Each logical qubit is path-encoded by a single photon in a pair of waveguides, with the parity of which waveguide represents $\ket{0}$ and $\ket{1}$ depending on the parity of the qubit index. 
    (b) A quantum circuit diagram depicting the logical representation of the operator performed by the QPGA in the first panel. The ``switch'' symbols between two-qubit operations indicate that the connectivity of the gates can be reconfigured without changing the physical chip architecture. Solid control dots indicate $\cz$, while open dots indicate $\czbar$.
    (c) A single unit cell within the lattice. The $\zeta, \xi, \theta, \phi$ phase shifters are continuously variable trainable parameters, while $\eta = 0, \frac{\pi}{2}$ determines the connectivity of the $\cz$ gates between neighboring qubits. The pink dots represent quantum emitters embedded a distance $a$ between two dichroic reflectors, depicted as blue and red rectangles, which selectively reflect light at frequencies $\omega$ and $\omega'$, respectively. The delay lines are matched in length to $\omega'$ and terminate in reflectors.
    (d) Four-level energy structure of the quantum emitters embedded in the waveguides.
    }
    \label{fig:lattice}
\end{figure*}

\section{Photonic quantum programmable gate arrays}
\label{sec:lattice_design}

The concept for a photonic quantum programmable gate array is shown in Fig. \ref{fig:lattice}a, and the equivalent logical quantum circuit is depicted in Fig. \ref{fig:lattice}b. The architecture consists of a set of waveguide pairs which each contain single photon pulses. A lattice of phase-modulated MZIs perform single-qubit rotations, and circulators, MZIs, and embedded four-level systems (4LS) collectively implement two-qubit controlled-$\sigma_z$ ($\cz$) gates between adjacent qubits. By choosing suitable phase shifter parameters, arbitrary multi-qubit quantum states and operators can be implemented from single-qubit and $\cz$ primitives within the lattice, as discussed in Section \ref{sec:implementing_quantum_operations}. In the following subsections, we discuss the mechanisms of each component of the architecture in greater detail.

\subsection{Single-qubit operations}
\label{sec:single_qubit_operations}

Qubits are implemented as temporally separated single photons, each injected into a pair of waveguides at a frequency $\omega$ and with a long pulse length $\tau \gg \omega^{-1}$. All physical gates within the device conserve photon occupancy within waveguide pairs. A chip designed to process $N$-qubit states has $2N$ number of waveguides, and the computational basis $\left\{\ket{0}, \ket{1} \right\}$ of each qubit is represented by the photon occupancy of the top and bottom waveguide in each pair, with parity alternating with qubit index as shown in Fig. \ref{fig:lattice}a. 

Single-qubit gates are implemented by phase-modulated MZIs. An MZI with four phase shifters in the configuration shown in the upper half of Fig. \ref{fig:lattice}c can apply any operation $U \in \mathrm{U}(2)$ to its inputs \cite{Pai2019MatrixDevices}, which suffices to implement arbitrary single-qubit gates. Assuming the photons are spectrally narrow about $\omega$ (see Appendix \ref{sec:interferometer_arbitrary_spectra} for a more complete treatment of arbitrary photon spectra), the transformation implemented by the MZI on the input modes takes the form:
\begin{align}
\begin{split}
\label{eq:mzi_four_phase_shifters}
    U(\zeta, \xi, \theta, \phi) &= R^\zeta_\xi H R^\theta_0 H R^\phi_0 \\
    &= \frac{1}{2}
    \begin{psmallmatrix} 
    e^{i \zeta} & 0 \\
    0 & e^{i \xi}
    \end{psmallmatrix}
    \begin{psmallmatrix} 
    1 & 1 \\
    1 & -1
    \end{psmallmatrix}
    \begin{psmallmatrix} 
    e^{i \theta} & 0 \\
    0 & 1
    \end{psmallmatrix}
    \begin{psmallmatrix} 
    1 & 1 \\
    1 & -1
    \end{psmallmatrix}
    \begin{psmallmatrix} 
    e^{i \phi} & 0 \\
    0 & 1
    \end{psmallmatrix} \\
    &= 
    \frac{1}{2} 
    \begin{bmatrix}
    e^{i (\zeta +\phi )} \left(e^{i \theta } + 1\right) & e^{i (\xi +\phi )} \left(e^{i \theta } - 1\right) \\
    e^{i \zeta } \left(e^{i \theta } - 1\right) & e^{i \xi } \left(e^{i \theta } + 1\right) \\
    \end{bmatrix},
\end{split}
\end{align}
where $H$ is the Hadamard operator\footnote{Whether to use $H=\frac{1}{\sqrt{2}} \left[\begin{smallmatrix} 1 & 1 \\ 1 & -1 \end{smallmatrix}\right]$ or $B=\frac{1}{\sqrt{2}} \left[\begin{smallmatrix} 1 & i \\ i & 1 \end{smallmatrix}\right]$ to represent the beamsplitter operation is somewhat a matter of convention, with classical optics tending to prefer the latter and quantum information often using the former. They are equivalent up to a phase shift of $\zeta, \theta$ by $\pi/2$. For simplicity, we choose to represent the beamsplitter operation here by $H$.} and $R^{\phi_1}_{\phi_2}$ denotes a phase shift of $\phi_1$ applied to the top waveguide and $\phi_2$ to the bottom. Here, and for the rest of this paper, successive matrices are left-multiplied to be consistent with circuit diagrams.

\subsection{Two-photon gates}
\label{sec:two_photon_gates}

In addition to arbitrary single-qubit gates, the QPGA needs to be able to implement two-qubit entangling operations in order to be a universal quantum device. This is accomplished by nonlinear interactions between two photons scattering off of a pair of quantum emitters embedded within the waveguides. The emitters could be implemented by quantum dots coupled to photonic crystal waveguides \cite{Kim2017HybridChip, Laucht2012ASource, Sollner2015DeterministicCircuits} or plasmonic nanowires \cite{Akimov2007GenerationDots}, diamond vacancy centers \cite{Radulaski2019NanodiamondDevices, Zhang2018StronglyDiamond, Babinec2010ASource}, or many other experimental setups. The scattering dynamics discussed in this section are based on the scheme described by Zheng et al. \cite{Zheng2013Waveguide-QED-basedComputation}, with the notable difference that spatial modes rather than momentum states form the computational basis for the physical qubits. In this section we show that this scattering process implements a logical $\cz$ operation up to local phase shifts.\footnote{Lattice cells of inverted parity (see Figure \ref{fig:lattice}a) actually implement $\czbar$, such that $\sigma_z$ is applied only to the $\ket{00}$ portion of the state, but the dynamics are the same, so for brevity we discuss only one parity here.}

Consider an arbitrary two-qubit logical input state $\ket{\Psi} = \alpha \ket{11} + \beta \ket{10} + \gamma \ket{01} + \delta \ket{00}$. The state consists of two photons superpositioned over two pairs of waveguides shown in Figure \ref{fig:lattice}. Define bosonic operators $\hat{a}^{[q]^\dagger}_{0,d}, \hat{a}^{[q]^\dagger}_{1,d}$ which create a photon with direction $d\in\{L,R\}$ in the $\ket{0}$ and $\ket{1}$ waveguide for qubit $q$, respectively. The corresponding two-photon physical input state $\ket{\psi}$ just before (1a, 1b) is:
\begin{align}
\begin{split}
\ket{\psi} 
&= \alpha \hat{a}^{[1]^\dagger}_{1,R} \hat{a}^{[2]^\dagger}_{1,R} \vac + \beta \hat{a}^{[1]^\dagger}_{1,R} \hat{a}^{[2]^\dagger}_{0,R} \vac \\
&+ \gamma \hat{a}^{[1]^\dagger}_{0,R} \hat{a}^{[2]^\dagger}_{1,R} \vac + \delta \hat{a}^{[1]^\dagger}_{0,R} \hat{a}^{[2]^\dagger}_{0,R} \vac,
\end{split}
\end{align}
where $\vac$ denotes the vacuum state (not to be confused with the computational $\ket{0}$ states). 

Consider the lower half of Figure \ref{fig:lattice}c. Two circulators at (1a, 1b) direct the $\ket{1}$ modal component of each photon into the waveguides at (2a, 2b). The photons pass through an MZI at (3) which has a transfer matrix\footnote{The reason for the $\pi/2$ phase shifters, which we'll temporarily refer to as $\gamma$, is that a photon traveling round-trip through this MZI undergoes a transformation $T^\transpose (\eta, \gamma) T(\eta,\gamma) = (R^\gamma_0 H R^\eta_0 H R^\gamma_0) (R^\gamma_0 H R^\eta_0 H R^\gamma_0)$. To conserve photon number within each waveguide pair, we require $T^\transpose T$ to be diagonal, so we must set $\gamma=\pi/2$.}:
\begin{equation}
\label{eq:eta_mzi_transfer_matrix}
    T(\eta) = R^{\pi/2}_0 H R^\eta_0 H R^{\pi/2}_0 = \frac{1}{2}
    \begin{pmatrix} 
    -e^{i\eta} - 1  & i e^{i\eta} - i \\
    i e^{i\eta} - i & e^{i\eta} + 1
    \end{pmatrix}
\end{equation}

Define bosonic operators $\hat{b}^{\text{top}^\dagger}_{d}$, $\hat{b}^{\text{bot}^\dagger}_{d}$, which create a photon in direction $d$ at (4a, 4b) respectively.\footnote{The $\hat{b}^{\text{top}^\dagger}$, $\hat{b}^{\text{bot}^\dag}$ notation was chosen to avoid confusion with the qubit indices or basis states $\hat{a}^{[q]^\dagger}_{\{0,1\}}$.} The MZI transfer matrix acts only on the $\ket{1}$ component of each photon, so we can relate the operators for each direction with:
\begin{equation}
\label{eq:beamsplitter_relations}
    \begin{pmatrix} \hat{a}^{[1]^\dagger}_{1,R} \\ \hat{a}^{[2]^\dagger}_{1,R} \end{pmatrix} = T(\eta) \begin{pmatrix} \hat{b}^{\text{top}^\dag}_{R} \\ \hat{b}^{\text{bot}^\dag}_{R} \end{pmatrix}, \quad \begin{pmatrix} \hat{a}^{[1]^\dagger}_{1,L} \\ \hat{a}^{[2]^\dagger}_{1,L} \end{pmatrix} = T^\transpose (\eta) \begin{pmatrix} \hat{b}^{\text{top}^\dag}_{L} \\ \hat{b}^{\text{bot}^\dag}_{L} \end{pmatrix},
\end{equation}
while $\hat{a}^{[q]^\dagger}_{0,d}$ are unaffected. Using the relations described in Eq. \ref{eq:beamsplitter_relations}, the input state after propagating through the MZI at (4a, 4b) is:
\begin{align}
\begin{split}
\label{eq:mixer_output_state}
\ket{\psi} &= 
e^{i \eta} \frac{\alpha}{2} \sin \eta \left( (\hat{b}^{\text{top}^\dag}_{R})^2 - (\hat{b}^{\text{bot}^\dag}_{R})^2 \right) \vac \\
&- e^{i \eta} \alpha \cos \eta \, \hat{b}^{\text{top}^\dag}_{R} \hat{b}^{\text{bot}^\dag}_{R} \vac \\
&- e^{i \eta/2} \left(\beta \cos \frac{\eta}{2} \hat{a}^{[2]^\dagger}_{0,R} + \gamma \sin \frac{\eta}{2} \hat{a}^{[1]^\dagger}_{0,R} \right) \hat{b}^{\text{top}^\dag}_{R} \vac \\
&- e^{i \eta/2} \left(\beta \sin \frac{\eta}{2} \hat{a}^{[2]^\dagger}_{0,R} - \gamma \cos \frac{\eta}{2} \hat{a}^{[1]^\dagger}_{0,R} \right) \hat{b}^{\text{bot}^\dag}_{R} \vac  \\
&+ \delta \hat{a}^{[1]^\dagger}_{0,R} \hat{a}^{[2]^\dagger}_{0,R} \vac.
\end{split}
\end{align}
The $\hat{b}^\dag$ photons propagate down the waveguides from (4a, 4b) until they interact with the embedded quantum emitters at (5a, 5b), while we assume the system acts trivially on the $\hat{a}^{[q]^\dag}_{0,R}$ photons.

We now consider the sections between (4a) to (6a) and (4b) to (6b). We will show that the two-photon state, upon passing through these sections, will gain a $\pi$ phase shift applied only to the first term of $\ket{\psi}$ in Eq. \ref{eq:mixer_output_state}, and thus a $\cz$ operation is implemented on the logical input state $\ket{\Psi}$. To show this, we first consider the dynamics of the photons in the section between sites (4a) to (6a) within a single isolated waveguide; the lower waveguide between (4b) and (6b) behaves identically. For simplicity, while we consider each waveguide in isolation, we drop the $\hat{b}^{\text{\{top,bot\}}}$ superscripts and omit the $\hat{a}^{[q]^\dagger}_{0,R}$ operators.

The regions of interest are shown in the middle of Fig. \ref{fig:lattice}c, which contain quantum emitters with the four-level energy structures shown in Fig. \ref{fig:lattice}d. The energy level of each state $\ket{i}$ is $\Omega_i$; we assume that $\Omega_4 - \Omega_3 = \Omega_2 - \Omega_1 = \omega$, and denote $\omega' \equiv \Omega_3 - \Omega_2$. The quantum emitters at (5a, 5b) are placed a distance $a$ between a pair of narrow-band filters, which are reflective at frequencies $\omega$ and $\omega'$, respectively, and transparent otherwise. Reflectors terminate the ends of the waveguides at (6a, 6b); the waveguides between the $\omega'$ filters and the reflectors form a delay line with a length which is a multiple of $\frac{2\pi}{\omega'}$. The real-space Hamiltonian that describes the coupling of such an atom to the waveguide without the filters is given by \cite{Rephaeli2013DissipationInvited, Shen2005CoherentWaveguides, Zheng2013Waveguide-QED-basedComputation}:
\begin{widetext}
\begin{multline}
\label{eq:hamiltonian}
\mathcal{H} = \frac{\hbar}{i} \int dx \left[ v_g \hat{b}_R^\dag (x) \frac{\partial}{\partial x} \hat{b}_R (x) - v_g \hat{b}_L^\dag (x) \frac{\partial}{\partial x} \hat{b}_L (x) + v_r \hat{c}^\dag (x) \frac{\partial}{\partial x} \hat{c}(x) \right] + \hbar \sum_{n=1}^{4} \Omega_n  \ketbra{n}{n} \\
+ \hbar \int dx \,\delta(x) \left[ \left(\sqrt{\frac{\Gamma v_g}{2}} \hat{b}_R^\dag (x) + \sqrt{\frac{\Gamma v_g}{2}} \hat{b}_L^\dag (x) + \sqrt{\Gamma' v_r} \hat{c}^\dag (x) \right) \left(\ketbra{1}{2} + \ketbra{3}{2} + \ketbra{3}{4} \right) + \text{H.c.} \right].
\end{multline}
\end{widetext}
Here, the first term describes the free waveguide dynamics, the second term describes the embedded four-level system shown in Fig. \ref{fig:lattice}d, and the third term is the interaction Hamiltonian. The decay rate into the waveguide is $\Gamma$, the coupling $\Gamma'$ describes the extrinsic loss of the excited states to degrees of freedom outside the waveguide, which is modeled as emission into a reservoir by the $\hat{c}^\dag, \hat{c}$ operators, and $v_g$ $\{v_r\}$ is the group velocity of the photons in the waveguide \{reservoir\}. The transition frequencies $\omega, \omega'$ obey $\left| \omega - \omega' \right| \gg \Gamma$. 

The scattering dynamics can be summarized by four steps occurring simultaneously in the top and bottom waveguides.
(1) Photon $A$ at frequency $\omega$ causes the atom, which is initialized in state $\ket{1}$, to partially transition from $\ket{1} \rightarrow \ket{3}$ with an amplitude of $\ket{3}$ corresponding to the photon occupancy in the waveguide. This emits an auxiliary photon $A'$ with frequency $\omega'$, which is reflected by one of the narrow-band mirrors and travels down the delay line. 
(2) While photon $A'$ is in the delay line, photon $B$, also at frequency $\omega$, is injected into the system. Interaction with the $\ket{1}$ component of the atomic states results in the transition $\ket{1}\rightarrow\ket{3}$ and releases an auxiliary photon $B'$ with frequency $\omega'$ down the delay line, while interaction with the $\ket{3}$ component imparts a $\pi$ phase shift onto $B$ and reflects it back into the waveguide.
(3) Photon $A'$ arrives back at the 4LS after traversing the delay line. By time reversal arguments, sending the output photon $A'$ back into the atom retrieves photon $A$, which exits the inner cell through its original waveguide.
(4) Photon $B'$ arrives back at the 4LS, retrieving photon $B$ as in step 3.

A conceptual animation depicting the two-photon scattering process in a QPGA cell can be found in the supplementary materials or at \url{github.com/fancompute/qpga}.

We now discuss each step in greater detail. Derivations of the reflection coefficients and output states can be found in Appendix \ref{sec:derivation_of_reflection_coefficients}.

\emph{Step 1.} 
At time $t=1$, photon $A$ with frequency $\omega$ and state $\ket{\psi^\text{in}_1} = \alpha_A \ket{\omega} + \beta_A \vac$ is incident on the 4LS, which is initialized to the state $\ket{1}$. From calculations detailed in Appendix \ref{sec:derivation_of_reflection_coefficients}, the output state is:
\begin{equation}
    \ket{\psi^\text{out}_1} = \alpha_A \left( r_{11} \ket{\omega} \otimes \ket{1} + r_{13} \ket{\omega'}\otimes \ket{3} \right) + \beta_A \vac \otimes \ket{1},
\end{equation}
where the amplitudes $r_{11}$ and $r_{13}$ are:
\begin{align}
    \label{eq:r11_main}
    r_{11} &= e^{2i \omega a} \frac
    {\Gamma' - \Gamma \left( e^{2i \omega' a} - e^{-2i \omega a} \right)}
    {-\Gamma' + \Gamma \left( e^{2i \omega' a} + e^{2i \omega a} - 2 \right)}, \\
    \label{eq:r13_main}
    r_{13} &= \frac
    {\Gamma \left( e^{2i\omega a} - 1 \right) \left( e^{2i\omega' a} - 1 \right)}
    {-\Gamma' + \Gamma \left( e^{2i \omega' a} + e^{2i \omega a} - 2 \right)}.
\end{align}
If the boundary condition that 
\begin{equation}
\label{eq:a_condition}
    a = \frac{n \pi}{\omega + \omega'}\text{ for some }n\in\mathbb{N}
\end{equation}
is satisfied, then $r_{11} = 0$ and $r_{13} = -1$, so $ \ket{\psi^\text{out}_1} = -\alpha_A \ket{\omega'}\otimes \ket{3} + \beta_A \vac \otimes \ket{1}$. Thus, the atom transitions from $\ket{1}\rightarrow\ket{3}$, stores the input photon, and releases an auxiliary $A'$ photon at frequency $\omega'$, which is reflected down the delay line.

\emph{Step 2.} 
At time $t=2$, photon $B$ with state $\ket{\psi^\text{in}_2} = \alpha_B \ket{\omega} + \beta_B \vac$ is incident on the 4LS. After scattering, the output state is:
\begin{align}
\begin{split}
\label{eq:psi2out}
    \ket{\psi^\text{out}_2} &= \alpha_B r_{11} \alpha_A r_{11} \ket{\omega}\otimes \ket{\omega} \otimes \ket{1} \\
    &+ \alpha_B r_{13} \alpha_A r_{11} \ket{\omega'} \otimes \ket{\omega} \otimes \ket{3} \\
    &+ \alpha_B R_3 \alpha_A r_{13} \ket{\omega} \otimes \ket{\omega'} \otimes \ket{3} \\
    &+ \alpha_B r_{11} \beta_A \ket{\omega} \otimes \vac \otimes \ket{1} \\
    &+ \alpha_B r_{13} \beta_A \ket{\omega'} \otimes \vac \otimes \ket{3} \\
    &+ \beta_B \vac \otimes \ket{\psi^\text{out}_1},
\end{split}
\end{align}
where the states are ordered as (photon $B$ $\otimes$ photon $A$ $\otimes$ atom), and where the reflection amplitude of the resonant $\ket{3}\rightarrow\ket{4}\rightarrow\ket{3}$ transition is derived in Appendix \ref{sec:derivation_of_reflection_coefficients} to be:
\begin{equation}
   R_3 = \frac{\Gamma' e^{2i\omega a} + \Gamma \left(1-e^{2i\omega a}\right)}
   {-\Gamma' - \Gamma \left(1-e^{2i\omega a}\right)}.
\end{equation}

As before, if the condition of Eq. \ref{eq:a_condition} is satisfied, then $R_3 = -1 = e^{i \pi}$, so photon $B$ gains a $\pi$ phase. For simplicity, in the rest of this section, we focus on the case where Eq. \ref{eq:a_condition} holds. Substituting the on-resonance coefficients of $r_{11} \rightarrow 0$, $r_{13} \rightarrow -1$, and $R_3 \rightarrow -1$ the output state at the end of step 2 is:
\begin{align}
\begin{split}
    \ket{\psi^\text{out}_2} &= \alpha_B \alpha_A \ket{\omega} \otimes \ket{\omega'} \otimes \ket{3} - \alpha_B \beta_A \ket{\omega'} \otimes \vac \otimes \ket{3} \\ 
    &- \beta_B \alpha_A \vac \otimes \ket{\omega'} \otimes \ket{3} + \beta_B \beta_A \vac \otimes \vac \otimes \ket{1}.
\end{split}
\end{align}

\emph{Step 3.} 
At time $t=3$, photon $A'$ has traveled down the delay line, which has a length which is a multiple of $\frac{2\pi}{\omega'}$, and is returning to the atom. Its frequency $\omega'$ is resonant with the $\ket{3}\leftrightarrow\ket{2}$ transition, and the reflection coefficients $r_{33}$ and $r_{31}$ have expressions which are identical to Eqs. \ref{eq:r11_main} and \ref{eq:r13_main}, respectively, except with $\omega, \omega'$ exchanged, such that when $a = \frac{n \pi}{\omega + \omega'}$, we have that $r_{33}=0$ and $r_{31}=-1$. 

The state of the returning $A'$ photon is $\ket{A'} = -\alpha_A \ket{\omega'} + \beta_A \vac$, and it only interacts with the $\ket{*}\otimes\ket{\omega'}\otimes\ket{3}$ components of the system state, mapping $\ket{*}\otimes\ket{\omega'}\otimes\ket{3} \mapsto -1 \cdot \ket{*}\otimes\ket{\omega}\otimes\ket{1}$. Therefore, the system state at the end of step 3 is:
\begin{align}
\begin{split}
    \ket{\psi^\text{out}_3} = &-\alpha_B \alpha_A \ket{\omega} \otimes \ket{\omega} \otimes \ket{1} - \alpha_B \beta_A \ket{\omega'} \otimes \vac \otimes \ket{3} \\
    &+ \beta_B \alpha_A \vac \otimes \ket{\omega} \otimes \ket{1} + \beta_B \beta_A \vac \otimes \vac \otimes \ket{1}.
\end{split}
\end{align}

\emph{Step 4.} 
At time $t=4$, photon $B'$ is returning to the 4LS from the delay line. The reflection coefficients are the same as in step 3, and photon only interacts nontrivially with the $\ket{\omega'}\otimes\ket{*}\otimes\ket{3}$ components of $\ket{\psi^\text{out}_3}$, so the final output state is:
\begin{align}
\begin{split}
\label{eq:psi4out}
    \ket{\psi^\text{out}_4} = &-\alpha_B \alpha_A \ket{\omega} \otimes \ket{\omega} \otimes \ket{1} + \alpha_B \beta_A \ket{\omega} \otimes \vac \otimes \ket{1} \\
    &+ \beta_B \alpha_A \vac \otimes \ket{\omega} \otimes \ket{1} + \beta_B \beta_A \vac \otimes \vac \otimes \ket{1}.
\end{split}
\end{align}

The state of the emitter at the end of the gate operation is restored to its original $\ket{1}$ state and is disentangled from photons $A$ and $B$, and the two-photon state acquires a $\pi$ phase shift only on the component corresponding to the presence of both $A$ and $B$. Thus, the gate operation in the computational basis of spatial modes is:
\begin{equation}
U = 
    \begin{psmallmatrix} 
    1 & 0 & 0 & 0 \\
    0 & 1 & 0 & 0 \\
    0 & 0 & 1 & 0 \\
    0 & 0 & 0 & -1
    \end{psmallmatrix},
\end{equation}
which is exactly the quantum controlled-$\sigma_z$ gate.

We now return to describing the evolution of the state where we left off at Eq. \ref{eq:mixer_output_state}. Using the $\ket{B} \otimes \ket{A} \otimes \ket{\text{4LS}}$ ordering from Eqs. \ref{eq:psi2out}-\ref{eq:psi4out}, we rewrite this equation to describe the states of the top and bottom photon-photon-4LS systems:
\begin{widetext}
\begin{multline}
\label{eq:psi_top}
\ket{\psi^\text{top}} = + e^{i \eta} \frac{\alpha}{2} \sin \eta \ket{\omega} \otimes \ket{\omega} \otimes \ket{1}
- \left(e^{i \eta} \alpha \cos \eta + e^{i \eta/2} \beta \cos \frac{\eta}{2} \right) \ket{\omega} \otimes \vac \otimes \ket{1} \\
- e^{i \eta/2} \gamma \sin \frac{\eta}{2} \vac \otimes \ket{\omega} \otimes \ket{1} 
+ \delta \vac \otimes \vac \otimes \ket{1}, 
\end{multline}
\begin{multline}
\label{eq:psi_bot}
\ket{\psi^\text{bot}} = - e^{i \eta} \frac{\alpha}{2} \sin \eta \ket{\omega} \otimes \ket{\omega} \otimes \ket{1} 
- e^{i \eta/2} \beta \cos \frac{\eta}{2} \ket{\omega} \otimes \vac \otimes \ket{1} \\
- \left(e^{i \eta} \alpha \cos \eta + e^{i \eta/2} \gamma \cos \frac{\eta}{2}\right) \vac \otimes \ket{\omega} \otimes \ket{1} 
+ \delta \vac \otimes \vac \otimes \ket{1}.
\end{multline}
\end{widetext}

The photons scatter off of the quantum emitters, producing ancillary photons which travel down the delay lines and back, which release the original photons, but with a $\pi$ phase shift applied to the $\ket{\omega} \otimes \ket{\omega} \otimes \ket{1}$ component of the state where both photons are present. Thus, the first term changes sign for each of Eqs. \ref{eq:psi_top} and \ref{eq:psi_bot}, and the output state when the photons finally return to the MZI in Figure \ref{fig:lattice}c, at (4a, 4b) is:
\begin{align}
\begin{split}
\ket{\psi} &= 
e^{i \eta} \frac{\alpha}{2} \sin \eta \left( -(\hat{b}^{\text{top}^\dag}_{L})^2 + (\hat{b}^{\text{bot}^\dag}_{L})^2 \right) \vac \\
&- e^{i \eta} \alpha \cos \eta \, \hat{b}^{\text{top}^\dag}_{L} \hat{b}^{\text{bot}^\dag}_{L} \vac\\
&- e^{i \eta/2} \left(\beta \cos \frac{\eta}{2} \hat{a}^{[2]^\dagger}_{0,L} + \gamma \sin \frac{\eta}{2} \hat{a}^{[1]^\dagger}_{0,L} \right) \hat{b}^{\text{top}^\dag}_{L} \vac \\
&- e^{i \eta/2} \left(\beta \sin \frac{\eta}{2} \hat{a}^{[2]^\dagger}_{0,L} - \gamma \cos \frac{\eta}{2} \hat{a}^{[1]^\dagger}_{0,L} \right) \hat{b}^{\text{bot}^\dag}_{L} \vac \\
&+ \delta  \hat{a}^{[1]^\dagger}_{0,R} \hat{a}^{[2]^\dagger}_{0,R} \vac,
\end{split}
\end{align}
where we assume that the photons described by the $\hat{a}^{[q]^\dagger}_{0,R}$ operators in Eq. \ref{eq:mixer_output_state} have been reflected and now travel in the $L$ direction.

Propagating this state through the MZI at (3) one last time using $\ket{\psi_\text{out}} = T^\transpose (\eta) \ket{\psi}$, we obtain the final output state at (2a, 2b):
\begin{align}
\begin{split}
\label{eq:psi_out}
\ket{\psi^\text{out}} &= 
e^{2i\eta} \alpha \cos (2\eta) \hat{a}^{[1]^\dagger}_{1,L} \hat{a}^{[2]^\dagger}_{1,L} \vac \\
&- e^{2i\eta} \frac{\alpha}{2} \sin(2\eta) \left((\hat{a}^{[1]^\dagger}_{1,L})^2 - (\hat{a}^{[2]^\dagger}_{1,L})^2 \right) \vac \\
& + e^{i \eta} \beta \hat{a}^{[1]^\dagger}_{1,L} \hat{a}^{[2]^\dagger}_{0,L} \vac \\
&+ e^{i \eta} \gamma \hat{a}^{[1]^\dagger}_{0,L} \hat{a}^{[2]^\dagger}_{1,L} \vac \\
&+ \delta \hat{a}^{[1]^\dagger}_{0,L} \hat{a}^{[2]^\dagger}_{0,L} \vac.
\end{split}
\end{align}
The output photons propagate to the circulators at (1a, 1b) and are reinjected back into their original waveguides. In order to preserve photon numbers between waveguide pairs, the second term in Eq. \ref{eq:psi_out} must be zero, since $(\hat{a}^{[1]^\dagger}_{1,L})^2$ and $(\hat{a}^{[2]^\dagger}_{1,L})^2$ correspond to injection of two photons into the same waveguide. This fact constrains $\eta$ to phase shifts which are multiples of $\frac{\pi}{2}$. 

We note the gate action of the entire system at $\eta=0$ is identity and the action at $\eta=\frac{\pi}{2}$ is the $\cz$ operation, up to a phase shift of $\frac{\pi}{2}$ which can be included in the $\zeta, \xi$ phase shifters at the subsequent column in the lattice:
\begin{align}
\label{eq:eta_gate}
U(\eta=0) &= 
    \begin{psmallmatrix} 
    1 & 0 & 0 & 0 \\
    0 & 1 & 0 & 0 \\
    0 & 0 & 1 & 0 \\
    0 & 0 & 0 & 1
    \end{psmallmatrix} = \id \\
U(\eta=\frac{\pi}{2}) &= 
    \begin{psmallmatrix} 
    1 & 0 & 0 & 0 \\
    0 & i & 0 & 0 \\
    0 & 0 & i & 0 \\
    0 & 0 & 0 & 1
    \end{psmallmatrix} = \left(R^{\pi/2}_0 \otimes R^0_{\pi/2}\right) \cz.
\end{align}

To summarize the results of this section, the photons are directed by circulators through an MZI and toward the scattering sites. Depending on the value of $\eta$, the four-level systems either interact with one ($\eta = 0$) or two ($\eta = \frac{\pi}{2}$) photons, and they impart a $\pi$ phase shift onto the two-photon component of the state they receive. The photons retrace their path and return to their original waveguides to be operated on by the next column of gates in the lattice.

\subsection{Fidelity and fault tolerance}

The calculations in the previous sections have shown that in an ideal case, our photonic architecture can perfectly implement arbitrary single-qubit operators and $\cz$. However, this makes some assumptions about the construction of the device. Namely, we assume that waveguides are lossless, that photons are injected with frequency $\omega$ and vanishing spectral width $\delta \omega \rightarrow 0$, and that the excited states of the scattering systems are lossless with $\Gamma' \rightarrow 0$, such that the Purcell enhancement factors are large, with $P = \Gamma / \Gamma' \rightarrow \infty$.

In reality, photons would have finite spectral width and the local emitters would have finite Purcell factors, meaning the QPGA would implement logical operators with fidelity below unity. As photons propagate through the imperfect gates implemented by the physical circuit, the errors will in general accumulate to render very deep circuits useless. However, this can be addressed by a variety of error-correcting methods. The errors which could occur in a physical implementation of this circuit can broadly be classified into three types: spectral unitary errors from the MZIs, photon loss, and depolarizing errors from the scattering sites.

MZIs acting on photons with finite spectral width (and dispersive effects in the waveguides) can reshape the photon pulse and transmit a portion of the pulse to the top and bottom waveguides which differs from the target amount. The photon is not lost to or entangled with the environment, so this error can be represented by a unitary operation $\tilde{U}$ with a characteristic error $\epsilon$ which acts as $\tilde{U} \ket{\psi^\text{in}} = \sqrt{1-\epsilon} \ket{\psi^\text{targ}} + \epsilon \ket{\psi^\text{targ}_{\perp}}$, where $\ket{\psi^\text{targ}_{\perp}}$ is some state orthogonal to the desired output state $\ket{\psi^\text{targ}}$. \cite{Koch2019SimulatingError} This error can, in principle, be trained around using the gradient-based circuit optimization approach discussed in Section \ref{sec:gradient-based-circuit-optimization}. However, as shown in Appendix \ref{sec:interferometer_arbitrary_spectra}, the fidelity of the MZIs can be quite high even for short pulses (a 1ns pulse has infidelity of $10^{-10}$), so the dominant source of error would come from the scattering operations.

Photon leakage from the waveguide or from spontaneous emission from the scattering sites can be corrected with only polynomial overhead by using concatenated coding \cite{Knill1996ConcatenatedCodes, Duan2004ScalableInteractions} or by using one of the Bose-Chaudhuri-Hocquenghem family of codes \cite{Grassl1997CodesChannel}.

The infidelity in the two-photon gates $\mathcal{F}^{-1} = 1-|\bra{\psi_A, \psi_B, \psi_\text{4LS}} \text{c}\sigma_{z}^{AB} \otimes \id^\text{4LS} \ket{\psi_A, \psi_B, 1}|^2$ introduced by finite excitation loss and spectral width results in a photon-photon-emitter state which is not fully entangled during operation nor fully disentangled at the end of the operation. If we trace out the degrees of freedom of the four-level system, we obtain a mixed two-photon output state which is the desired output state, but with a probability $p=\mathcal{F}^{-1}$ of applying a second $\sigma_z$ operation which undoes the original gate action. This corresponds to the well-studied quantum depolarizing error model \cite{Knill2002IntroductionCorrection, Gottesman2009AnComputation, Nielsen2010QuantumInformation}, which describes quantum gates as being faulty by randomly applying Pauli operators with some effective error probability per gate (EPG). \cite{Knill2005QuantumDevices, Nielsen2010QuantumInformation} Fault tolerance requires an EPG below a certain threshold $p_\text{th}$, usually estimated as $p_\text{th} \approx 10^{-4}$ \cite{Gottesman1997StabilizerCorrection, Preskill1997Fault-tolerantComputation}, but for some architectures and scenarios as high as $p_\text{th} \approx 10^{-2}$ \cite{Knill2005QuantumDevices}. In this system, the EPG approaches zero as the Purcell factor tends to infinity, with $P=40$ yielding a $6\%$ infidelity. \cite{Zheng2013Waveguide-QED-basedComputation} Recent experimental progress in waveguide-cavity systems have demonstrated very large Purcell factors ($P>80$) \cite{Bracher2017SelectiveCenter}, and further improvement can be expected from rapid development of waveguide technologies, so large-scale QPGAs may be feasible to implement in the foreseeable future.

\section{Exact quantum state and operator preparation}
\label{sec:implementing_quantum_operations}

Having established how the design presented in Section \ref{sec:lattice_design} acts on physical photonic qubits, we now discuss how the idealized logical model of the device can be programmed to prepare quantum states and to implement quantum operators. We assume no error in the device here and describe algorithms to implement the desired actions with perfect fidelity, albeit sometimes using circuits of great depth. In reality, finite device errors may make the more compact approximate circuit decompositions discussed in Section \ref{sec:gradient-based-circuit-optimization} more relevant than the exact decompositions presented in this section.

\subsection{Universality of the design}

The MZIs in the lattice can implement any single-qubit gate by parameterizing it through the $\zeta, \xi, \theta, \phi$ phase shifts. The nonlinear interactions between waveguide pairs implements $\cz$, which can be used in conjunction with $H$ to implement a controlled-NOT ($\cx$) gate\footnote{Due to the nearest-neighbor connectivity of the architecture, $\cx$ between non-adjacent qubits must be implemented with a sequence of SWAP gates, which can in turn be implemented using three $\cx$ gates. \cite{Nielsen2010QuantumInformation}} as $\cx = (\id \otimes H) \cz (\id \otimes H)$. \cite{Nielsen2010QuantumInformation} Since the set of single-qubit operations and $\cx$ gate comprises a universal gate set \cite{Barenco1995ElementaryComputation}, the device is universal, such that a sufficient number of layers can be used to implement an arbitrary multi-qubit gate. 

Phase shifter parameters which implement various common single- and two-qubit quantum gates are detailed in Appendix \ref{sec:gate_implementations}. Notably, two-qubit gates can have differing $\cz$ parities, meaning that some require an even or odd number of successive $\cz$ gates to implement. This would be problematic in an architecture with fixed $\cz$ connectivity, as aligning circuit elements within a fixed lattice would be impossible; this necessitates a mechanism such as the $\eta$-shifted MZI described by Eq. \ref{eq:eta_mzi_transfer_matrix} which can toggle the gate action between qubits.

\subsection{State preparation}
\label{sec:state_preparation}

Arbitrary quantum states can be prepared on a lattice with nearest-neighbor connectivity using a circuit based on Ref. \cite{Kaye2004QuantumStates} consisting of a sequence of multi-controlled single-qubit rotations. Although the general worst-case complexity of this algorithm is $\mathcal{O}(n^2 2^n)$, an important class of quantum states, including Dicke states \cite{Bartschi2019DeterministicStates} and general symmetric states \cite{Kaye2004QuantumStates}, can be efficiently prepared using such a lattice with a depth which is polynomial in the number of qubits.

Suppose we have a state $\ket{\psi}=\sum_{q\in\{0,1\}^n} \alpha_q \ket{q}$ with $\alpha_q \in \mathbb{C}$ which we would like to prepare. Let $\xi_x$ for $x\in\{0,1\}^k$ and $1 \le k \le n$ denote the projection of $\ket{\psi}$ onto the computational basis vector $\ket{x}$, tracing over all qubits subsequent to $k$:
\begin{equation}
    \xi_x = \sum_{x'\in\{0,1\}^{n-k}} \braket{x, x' | \psi}.
\end{equation}

For each string $x$ of length $k$, define a $k$-ly controlled single-qubit rotation operator $U_{x_1 \cdots x_k}$ acting on qubit $k+1$ which maps:
\begin{align}
\begin{split}
    U_{x_1 \cdots x_k} \ket{x_1 \cdots x_k} \ket{0} &= \frac{\xi_{x_1 \cdots x_k 0}}{\xi_{x_1 \cdots x_k}} \ket{x_1 \cdots x_k}\ket{0} \\
    &+ \frac{\xi_{x_1 \cdots x_k 1}}{\xi_{x_1 \cdots x_k}} \ket{x_1 \cdots x_k}\ket{1}.
\end{split}
\end{align}
Each $k$-controlled operation can be implemented on the nearest-neighbor architecture of the lattice with $\mathcal{O}(k^2)$ depth in the lattice using the implementation depicted in Figure 4.10 of Ref. \cite{Nielsen2010QuantumInformation}.

\begin{figure}[ht]
\resizebox{\columnwidth}{!}{
\mbox{
\scriptsize
\Qcircuit @C=1em @R=1em {
&\lstick{\ket{0}} & \gate{U} & \ctrlo{1}      & \ctrl{1}       & \ctrlo{1} 		 & \ctrlo{1} 		 & \ctrl{1} 		& \ctrl{1} 		& \qw & \cdots & & \qw & \ctrl{5} 				& \ctrl{5} 				& \qw \\
&\lstick{\ket{0}} & \qw          & \gate{U_{0}}   & \gate{U_{1}} & \ctrlo{1} 		 & \ctrl{1} 		 & \ctrlo{1} 		& \ctrl{1} 		& \qw & \cdots & & \qw & \ctrl{4} 				& \ctrl{4} 				& \qw \\
&\lstick{\ket{0}} & \qw          & \qw            & \qw            & \gate{U_{00}} 	 & \gate{U_{01}} 	 & \gate{U_{10}}	& \gate{U_{11}} & \qw & \cdots & & \qw & \ctrl{3} 				& \ctrl{3} 				& \qw \\
&\lstick{}        & \vdots       &                &                &     			 &     				 &     				&     			&     & \vdots & &     &          				&          				&     \\
&\lstick{}        &              &                &                &     			 &     				 &     				&     			&     &        & &     &          				&          				&     \\
&\lstick{\ket{0}} & \qw          & \qw            & \qw            & \qw 			 & \qw 				 & \qw 				& \qw 			& \qw & \cdots & & \qw & \ctrlo{1} 				& \ctrl{1} 				& \qw \\
&\lstick{\ket{0}} & \qw          & \qw            & \qw            & \qw 			 & \qw 				 & \qw 				& \qw 			& \qw & \cdots & & \qw & \gate{U_{11\cdots 0}} & \gate{U_{11\cdots 1}} & \qw \\
}
}
}
\caption{State preparation algorithm to map $\ket{0}^{\otimes n} \mapsto \ket{\psi}$}
\label{fig:state_prep_algorithm}
\end{figure}
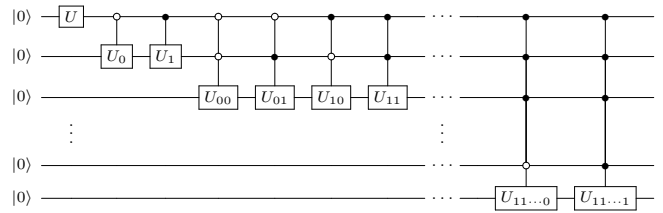

The brute-force algorithm for preparing $\ket{\psi}$ is the application of $2^n$ of these operations, as shown in the circuit diagram of Figure \ref{fig:state_prep_algorithm}. 

It can be shown by induction that after the first $k$ rotations, the resulting state is $\sum_{x_1 \cdots x_k \in\{0,1\}^k} \xi_{x_1 \cdots x_k} \ket{x_1 \cdots x_k}$, so after all $2^n$ operations, the output state is:
\begin{equation}
    \sum_{x_1 \cdots x_n \in\{0,1\}^n} \xi_{x_1 \cdots x_n} \ket{x_1 \cdots x_n} = \sum_{q\in\{0,1\}^n} \alpha_q \ket{q} = \ket{\psi}.
\end{equation}

Although this algorithm is not efficient for arbitrary quantum states, it is capable of efficiently preparing many interesting and important states. For example, an $n$-qubit GHZ state can be prepared on a nearest-neighbor lattice using $\mathcal{O}(n)$ by setting $U=H$, applying singly-controlled $\cx$ between successive qubits, and discarding all other $U_{x_1 x_2 \cdots x_k}$ operators.

\subsection{Implementation of general quantum operators}

Arbitrary $U(2^n)$ operations can be exactly implemented on the lattice using a nullification algorithm similar to the decomposition routines for classical optical meshes presented in Refs. \cite{Reck1994ExperimentalOperator, Clements2016AnInterferometers}. A more in-depth treatment of this problem can be found in Ref. \cite{Mottonen2005DecompositionsGates}.

In linear algebra, QR factorization decomposes any unitary matrix as $U=QR$, where $R$ is diagonal and unitary and $Q$ is a product of two-level Givens rotations \cite{Givens1958ComputationForm, Mottonen2005DecompositionsGates}, which are operations acting trivially on all but two basis vectors $\ket{m}, \ket{n}$: 
\begin{align}
\begin{split}
G_{m,n}(\theta, \phi) 
&= e^{i\phi} \cos \theta \ketbra{m}{m} - \sin \theta \ketbra{m}{n} \\
&+ \,e^{i\phi} \sin \theta \ketbra{n}{m} \; + \cos \theta \ketbra{n}{n}.
\end{split}
\end{align}

For any unitary matrix $U$, there exist values of $\theta,\phi$ which ``nullifies'' a target element in row $m$ or $n$ of $U$. \cite{Clements2016AnInterferometers} Let $G_{m,n}^{j}$ denote the Givens rotation to nullify the element of $U$ in row $m$, column $j$ against the element in row $n$, column $j$. It can be shown \cite{Mottonen2005DecompositionsGates} that after applying $\mathcal{O}(4^n)$ Givens rotations, we obtain an identity matrix:
\begin{equation}
\left[ \prod_{j=1}^{2^{n}-1} \prod_{m=j+1}^{2^n} G_{m,m-1}^{2^n-j} \right] U = \id.
\end{equation}

The operations $G_{m,m-1}^{2^n-j}$ do not correspond to any standard quantum gates, but if the basis vectors are permuted to be ordered in the reflected binary code \cite{Weisstein1999GrayCode}, then the Givens rotations between adjacent vectors $\ket{m}, \ket{m-1}$ can be written as a product of $(n-1)$-ly controlled single-qubit rotations \cite{Vartiainen2004EfficientGates}, each of which can be performed with a lattice depth of $\mathcal{O}(n^2)$. Thus, the target operator $U$ can be implemented as:
\begin{equation}
U = \prod_{j=1}^{2^{n}-1} \prod_{m=j}^{2^{n}-j} G_{\gamma(2^n-m+1),\gamma(2^n-m)}^{\gamma(j)^\dag},
\end{equation}
where $\gamma(j)$ denotes the index $j$ in reflected binary ordering. The permutation for each of the $\mathcal{O}(4^n)$ Givens rotations requires $\mathcal{O}(n^3)$ $\cx$ gates, so the worst-case complexity is $\mathcal{O}(n^3 4^n)$. 

As with state preparation, although implementing the most general quantum operators is hard, many important quantum operators, such as the quantum Fourier transform, may be efficiently implemented using a lattice of polynomial depth.

\section{Gradient-based circuit optimization}
\label{sec:gradient-based-circuit-optimization}

In the previous section we discussed preparation of arbitrary quantum states or operators by obtaining appropriate phase shifter values to implement an exact decomposition of the desired operation using only single-qubit and nearest-neighbor $\cz$ gates. In this section, we demonstrate a method, building on our previous work for classical MZI networks \cite{Pai2019MatrixDevices, Williamson2020ReprogrammableNetworks} and on work for continuous-variable quantum neural networks \cite{Arrazola2018MachineComputers}, of automatically discovering high-fidelity approximate decompositions of a target operator using a gradient-based optimization approach. As shown in Section \ref{sec:compactness_analysis}, these ``learned'' implementations of quantum operators are often far more compact than an explicit decomposition, allowing for lattices with a fraction of the physical depth.

Let $U_{il} = U (\zeta_{il},\xi_{il},\theta_{il},\phi_{il})$ denote the operation described by Eq. \ref{eq:mzi_four_phase_shifters} acting on qubit $i$ performed by a single MZI in layer $l$ of the lattice. Each layer of the lattice refers to the column of MZIs implementing $U_{il}$ and a subsequent column of $\text{c}\sigma_z^{i,j}$ gates between qubits $i$ and $j$.

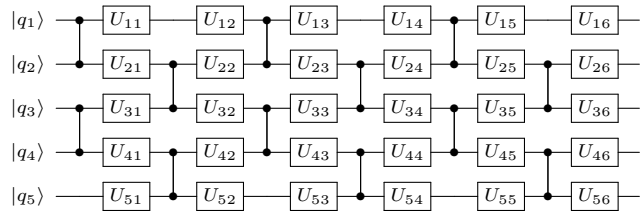
\begin{figure}[ht]
\resizebox{\columnwidth}{!}{
\mbox{
\scriptsize
\Qcircuit@C=1.0em@R=.75em{
&&\lstick{\ket{q_1}}	&\ctrl{1} 	&\gate{U_{11}}	&\qw		&\gate{U_{12}}	&\ctrl{1} 	&\gate{U_{13}}	&\qw		&\gate{U_{14}}	&\ctrl{1} 	&\gate{U_{15}}	&\qw		&\gate{U_{16}}	&\qw		\\
&&\lstick{\ket{q_2}}	&\ctrl{-1}	&\gate{U_{21}}	&\ctrl{1} 	&\gate{U_{22}}	&\ctrl{-1}	&\gate{U_{23}}	&\ctrl{1} 	&\gate{U_{24}}	&\ctrl{-1}	&\gate{U_{25}}	&\ctrl{1} 	&\gate{U_{26}}	&\qw		\\
&&\lstick{\ket{q_3}}	&\ctrl{1} 	&\gate{U_{31}}	&\ctrl{-1}	&\gate{U_{32}}	&\ctrl{1} 	&\gate{U_{33}}	&\ctrl{-1}	&\gate{U_{34}}	&\ctrl{1} 	&\gate{U_{35}}	&\ctrl{-1}	&\gate{U_{36}}	&\qw		\\
&&\lstick{\ket{q_4}}	&\ctrl{-1}	&\gate{U_{41}}	&\ctrl{1} 	&\gate{U_{42}}	&\ctrl{-1}	&\gate{U_{43}}	&\ctrl{1} 	&\gate{U_{44}}	&\ctrl{-1}	&\gate{U_{45}}	&\ctrl{1} 	&\gate{U_{46}}	&\qw 		\\
&&\lstick{\ket{q_5}}	&\qw		&\gate{U_{51}}	&\ctrl{-1}	&\gate{U_{52}}	&\qw		&\gate{U_{53}}	&\ctrl{-1}	&\gate{U_{54}}	&\qw		&\gate{U_{55}}	&\ctrl{-1}	&\gate{U_{56}}	&\qw 
}
}
}
\caption{Fixed connectivity scheme employed in training. The $\cz$ operators in odd columns are implicitly constructed by embedding $\sigma_x$ operations before and after physical $\czbar$ gates.}
\label{fig:fixed_connectivity}
\end{figure}

Because the strength of the $\cz$ interaction is not a continuous variational parameter (since the only valid settings are $\eta=0$ (off) or $\eta=\frac{\pi}{2}$ (on), as discussed in Section \ref{sec:two_photon_gates}), in our numerical experiments, we employ a checkerboard-style connectivity where half of the $\cz$ gates are disconnected, as shown in Figure \ref{fig:fixed_connectivity}. In a given layer, the $\cz$ gates are applied to each pair of adjacent qubits with an offset determined by the parity of the layer index. Additionally, we implicitly embed logical $\sigma_x$ gates in the single-qubit operators preceding and following two-qubit gates in odd layers, such that $U_{i, n} \mapsto U_{i, n} \sigma_x$ and $U_{i, n+1} \mapsto \sigma_x U_{i, n+1}$ for odd $n$; this transforms the $\czbar$ gates applied in odd layers into $\cz$ gates without adding to the depth of the lattice.

The operation performed on an $N$-qubit quantum state by a lattice of depth $L$ with this connectivity scheme is given by:
\begin{equation}
\mathcal{U}_{\vec{\Theta}} = \prod_{l=1}^{L} \left[ \bigotimes_{i \in \mathcal{C}(l)} \text{c}\sigma_z^{i,i+1} \cdot \bigotimes_{i=1}^N U (\zeta_{il},\xi_{il},\theta_{il},\phi_{il}) \right],
\end{equation}
where $\vec \Theta$ denotes all free parameters $\{\zeta_{il},\xi_{il},\theta_{il},\phi_{il}\}$ in the lattice, where the set of $\cz$ connections is $\mathcal{C}(l) = \{1,3,5,\cdots ,2 \ceil{\frac{N}{2}} - 1\}$ [$\mathcal{C}(l) = \{2,4,6,\cdots ,2 \floor{\frac{N}{2}}\}$] for odd [even] $l$, and where left-multiplication and padding with identity are implicit. 

Let $\mathcal{F}(\tilde{\psi}, \psi) = \abs{\braket{\tilde{\psi} | \psi}}^2$ denote the fidelity between states $\ket{\tilde\psi}$ and $\ket{\psi}$. To implement a target operator $\hat{U}$, the optimization routine finds a set of parameters $\vec \Theta$ which maximizes the average fidelity $\mathcal{F}= |\bra{\psi_\text{in}} \mathcal{U}_{\vec{\Theta}}^{\dagger} \hat{U} \ket{\psi_\text{in}}|^2$ over a ``training set" of input states $\{\psi_\text{in}\}$. The algorithm computes the gradient $\nabla_{\vec \Theta} \mathcal{F}$ of the fidelity over the training states with respect to the phase shift parameters and iteratively updates $\vec \Theta$ by a step size $\eta$ as $\vec \Theta \mapsto \vec \Theta + \eta \nabla_{\vec \Theta} \mathcal{F}$ over the course of the training. In the case of operator implementation, $\{\psi_\text{in}\}$ are an ensemble of uniformly randomly sampled state vectors, while for state preparation, $\{\psi_\text{in}\} = \{\ket{0}^{\otimes N}\}$.

\subsection{Simulations}
\label{sec:simulations}

In the following subsections we present a series of numerical experiments in which a simulated logical model of a QPGA is trained to implement a variety of quantum states and operators. The numerical model was programmed using a custom backend built with $\mathtt{TensorFlow}$ \cite{Abadi2016TensorFlow:Learning}, and the source code for all experiments in this paper is available at \url{github.com/fancompute/qpga}. 

For operator preparation simulations, we generate the training set $\{\psi_\text{in}\}$ of random $n$-qubit state vectors by randomly choosing $2^n$ component magnitudes uniformly between $[0,1)$, then renormalizing the state vector and assigning each component a random phase between $[0,2\pi)$. The number of training samples is empirically chosen, but always greatly exceeds $2^n$. The corresponding target output states are produced by running the input states through an explicitly constructed quantum circuit simulated using the $\mathtt{SQUANCH}$ Python framework \cite{Bartlett2018AChannels}. For state preparation simulations, the training set is simply the zero state input $\ket{0}^{\otimes N}$ and the corresponding single output state is directly compared against the target state.

For all simulations, we used the checkerboard connectivity scheme described in the previous section. We initialized all $\zeta, \xi, \theta, \phi$ phase shifters uniformly from $[0,2\pi)$, optimized the gate array using the Adam optimizer \cite{Kingma2014Adam:Optimization} with learning rate annealing, and performed the training on an NVIDIA Tesla K80 GPU.

\subsubsection{GHZ state preparation}

Greenberger–Horne–Zeilinger (GHZ) states \cite{Greenberger2007GoingTheorem} are maximally-entangled multi-qubit states of the form $\frac{1}{\sqrt 2}(\ket{00\cdots0} + \ket{11\cdots1})$ and have important applications in quantum information and quantum cryptography. \cite{Hillery1999QuantumSharing}

\begin{figure}[ht]
    \centering
    \includegraphics[width=\columnwidth]{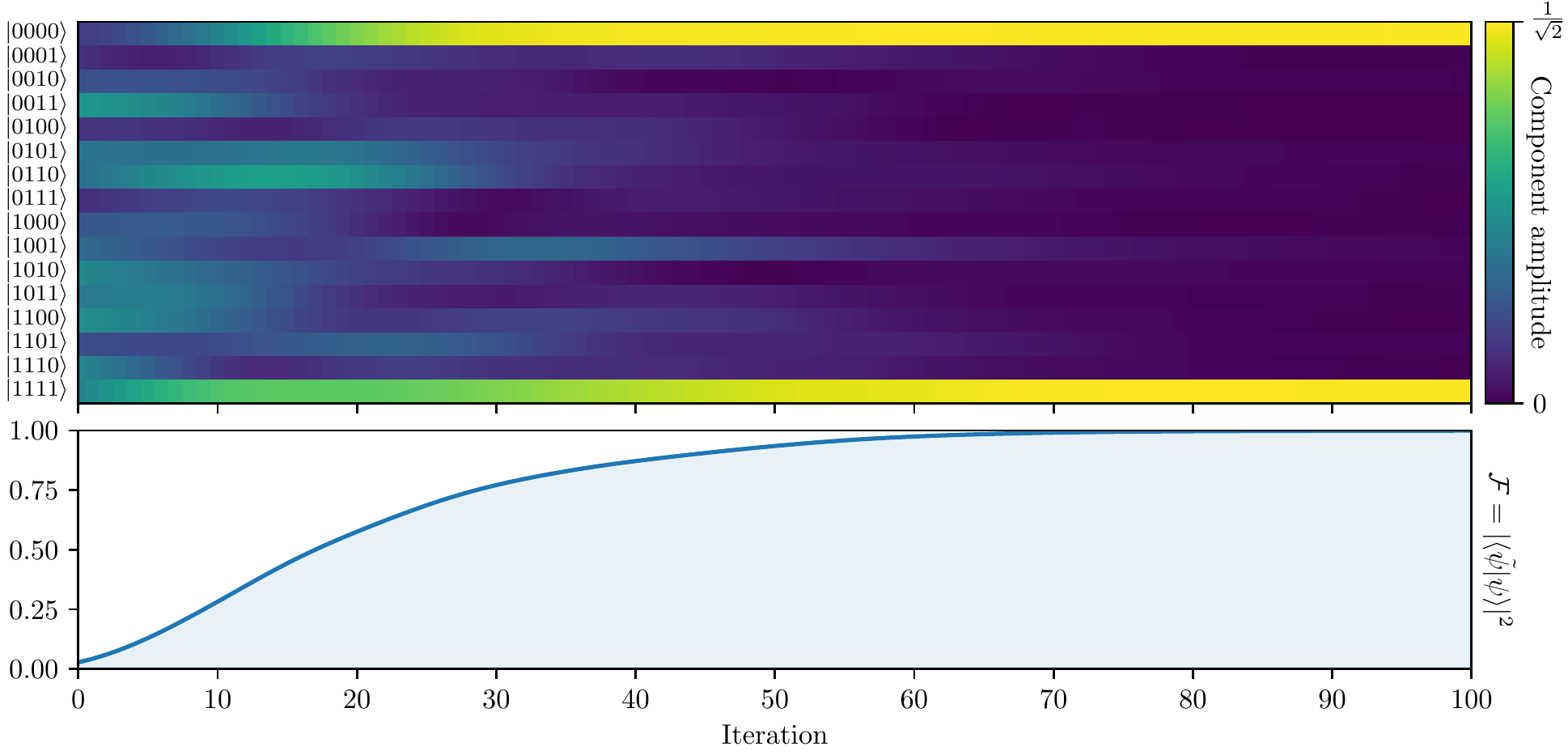}
    \caption{Optimization of a quantum circuit to prepare a four-qubit GHZ state. (Top) Evolution of the output state $\ket{\tilde \psi}$ over the course of training. The vertical axis represents the magnitude of the projection $\braket{\tilde{\psi} | b_j}$ of the output state onto each computational basis state $\ket{b_j}$. (Bottom) Fidelity between the output state and target state over the course of training, reaching a maximum value of $\mathcal{F} \approx 99.94\%$. The shared horizontal axes on both panes indicates iterations during training.}
    \label{fig:ghz_state_prep}
\end{figure}

Figure \ref{fig:ghz_state_prep} shows the optimization progress of a four-qubit GHZ state. We simulated a small four-qubit QPGA with a fixed depth of 20 layers. For visualization purposes, we used a low learning rate and only displayed the first 100 iterations of training. Using a deeper lattice with longer training, arbitrarily high fidelities can be reached.

The stochastic nature of the initialization and optimization routines means that the training converges non-deterministically. Shallower circuits have fewer variational parameters to optimize and fewer layers to allow entanglement to propagate between nearest-neighbor qubits, which can result in a final fidelity which is far from unity. Deeper circuits have more parameters to optimize but require greater computational resources to simulate (and experimentally would have more pronounced physical errors if this were being considered). The number of layers in the circuit was empirically chosen to be a small depth which would consistently reach $\mathcal{F}\approx 1$.

Due to the uniform initialization of the phase shifters in the lattice, the model initially outputs a random, non-maximally entangled quantum state lacking any apparent structure. As the optimization routine proceeds, the lattice produces states which have increasingly large $\ket{0000}$ and $\ket{1111}$ components, with the relative phase between these components approaching $0$, while the other components of the output state have vanishing amplitudes. After 100 iterations, the model generates a state matching the target state with $99.94\%$ fidelity.

\subsubsection{Random state preparation}

\begin{figure}[ht]
    \centering
    \includegraphics[width=\columnwidth]{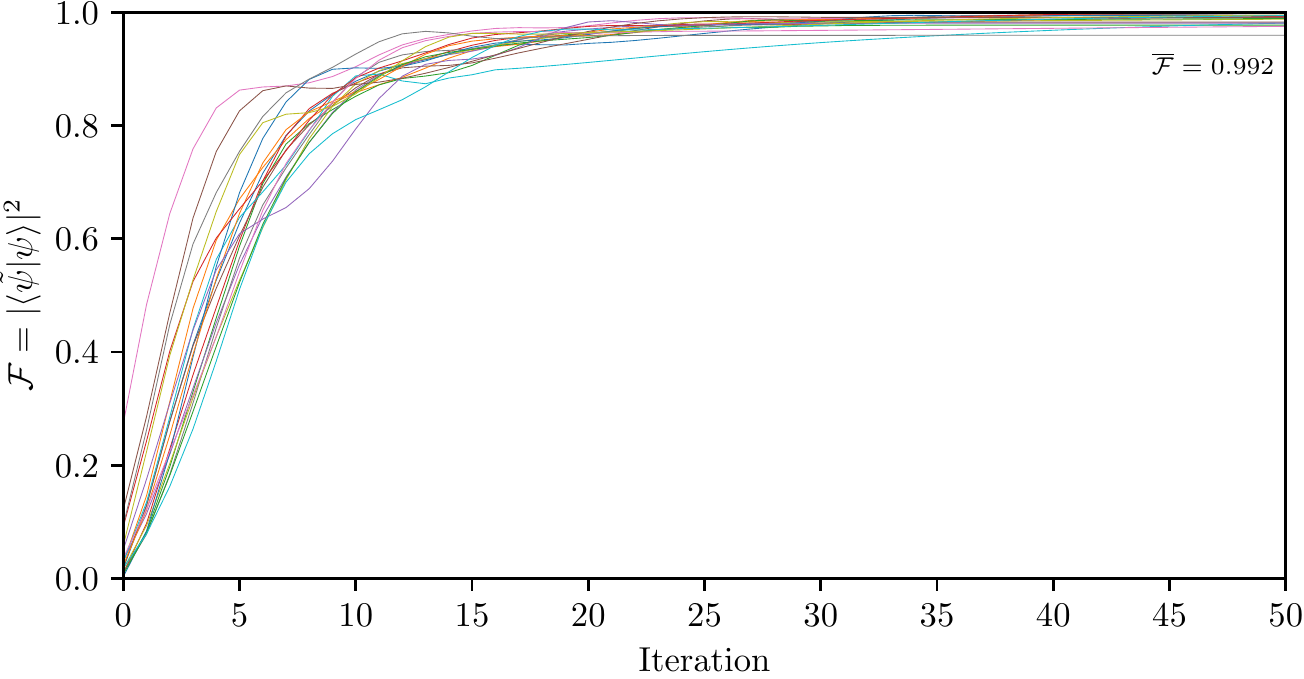}
    \caption{Training a 20-layer QPGA to prepare an ensemble of randomly sampled four-qubit states. Fidelities between the output and target states are shown over the course of each optimization. The average fidelity at the end of training is $\overline{\mathcal{F}} = 99.2 \%$.}
    \label{fig:random_state_prep}
\end{figure}

As discussed in Section \ref{sec:state_preparation}, states with certain structures and symmetries are easier to prepare than general quantum states. To demonstrate the generality of the gradient-based circuit optimization routine, we use it to prepare a sample of random quantum states.\footnote{While the gradient-based circuit decomposition method will not bypass the exponential complexity of approximating general quantum states and operators (see Ref. \cite{Nielsen2010QuantumInformation}, section 4.5.4), it is still informative to show that the method can implement states without specific structure.} The states are generated by choosing $2^n$ component magnitudes and phases uniformly, as described at the beginning of Section \ref{sec:simulations}. We choose $n=4$ qubits and fix a depth of 20 layers; the fidelities between the output states and target states over the course of training is shown in Figure \ref{fig:random_state_prep}. The results show that a QPGA of this depth is sufficient to create an arbitrary 4-qubit state with high fidelity.

\begin{figure*}[t]
    \centering
    \includegraphics[width=\textwidth]{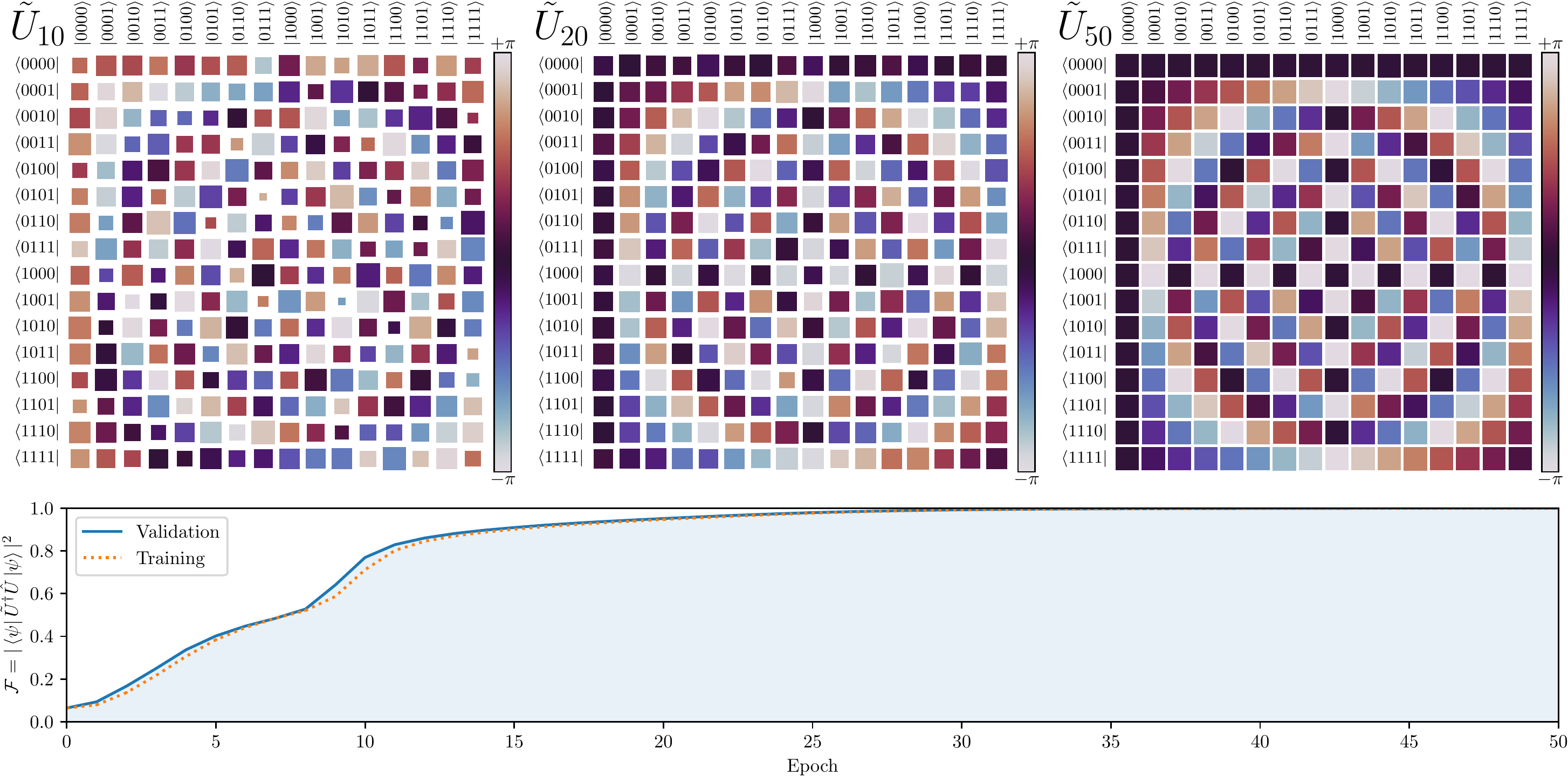}
    \caption{Optimization of a 20-layer QPGA to prepare a quantum Fourier transform on the four input qubits. (Top) The operators $\tilde{U}_i$ implemented by the QPGA after $i$ training epochs. Each square array represents the magnitude (relative to the maximum element) and phase of the projection of the operator onto the lexicographically-ordered computational basis states, encoded in the respective size and hue of the squares. The final $\tilde{U}_{50}$ is visually indistinguishable from $\hat{U}$. (Bottom) Fidelity between the implemented and target operator over the course of training. The final fidelity is $\mathcal{F}=99.94\%$. An animated version of this figure showing the evolution of the implemented operator over the course of training can be found in the supplementary materials or at \url{github.com/fancompute/qpga}.}
    \label{fig:qft}
\end{figure*}

\subsubsection{Quantum Fourier transform}

The quantum Fourier transform is an important operator which plays a key role in many quantum computing algorithms, especially the eigenvalue estimation routine. \cite{Nielsen2010QuantumInformation} The quantum Fourier transform operating on $n$ qubits takes the form:
\begin{equation}
    U_\text{QFT}=\frac{1}{n} \sum_{j=0}^{n-1} \sum_{k=0}^{n-1} e^{2\pi i j k / n} \ketbra{j}{k}.
\end{equation}

For this simulation, we compare the trainable circuit against the exact circuit implementation of the QFT, which has a complexity of $\mathcal{O}(n^2)$ (although the QFT can be approximated to within an inverse polynomial in $n$ using only $\mathcal{O}(n \log n)$ gates \cite{Hales2000ImprovedApplications}).

Figure \ref{fig:qft} shows the optimization of a QPGA to implement a quantum Fourier transform on four input qubits. The explicit decomposition of the QFT circuit requires 57 layers\footnote{We train against the explicit circuit provided in Ref. \cite{Nielsen2010QuantumInformation}, Fig. 5.1, but additionally add $\floor{\frac{n}{2}}$ SWAP gates, since the output qubits in the Fourier basis are otherwise in reverse order.}, but a trained QGPA with only 20 layers achieves a near-unity fidelity of $\mathcal{F}=99.94\%$.

\subsubsection{Circuit compactness analysis}
\label{sec:compactness_analysis}

\begin{figure}[ht]
    \centering
    \includegraphics[width=\columnwidth]{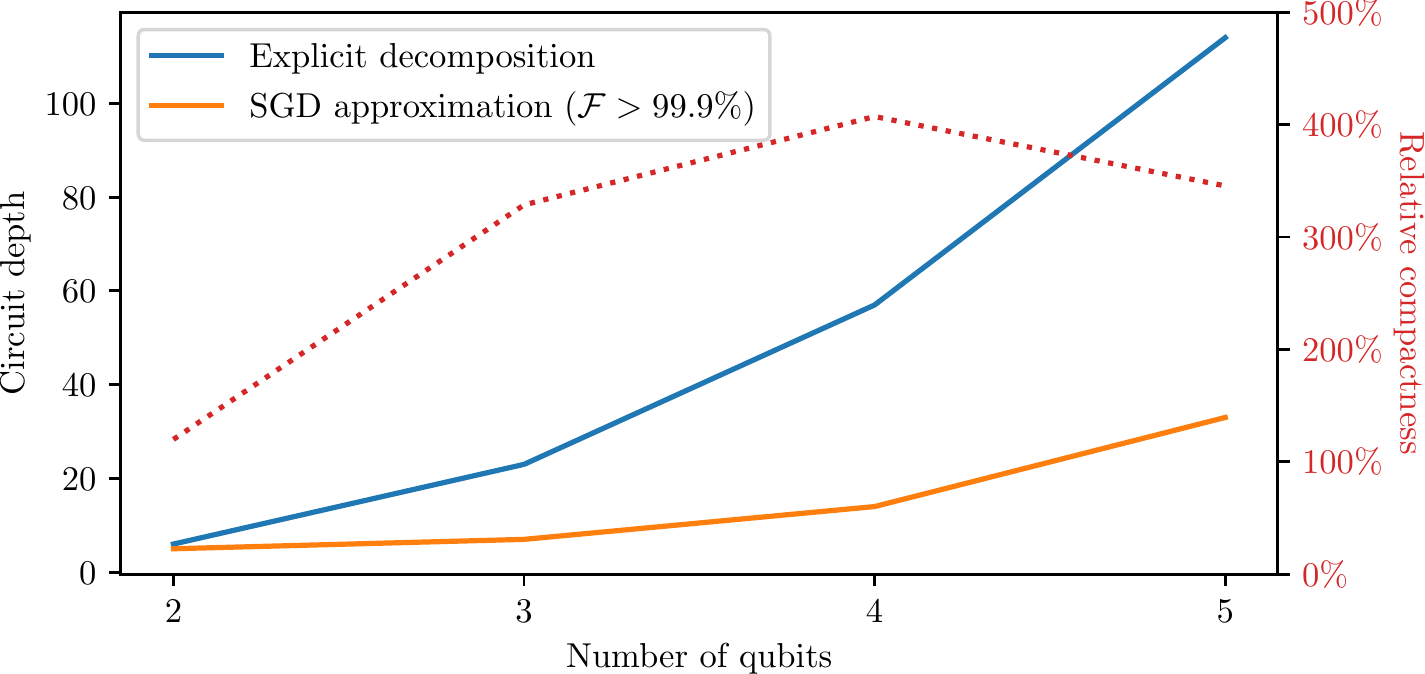}
    \caption{Required circuit depths to implement a quantum Fourier transform for a range of qubit numbers using explicit decomposition and using gradient-based decomposition which achieves a fidelity above $99.9\%$. The approximate decompositions are significantly more compact than the explicitly constructed circuits.}
    \label{fig:circuit_depth}
\end{figure}

In the previous sections, we have shown that gradient-based circuit optimization can produce high-fidelity operators which are significantly more compact than their explicitly-decomposed counterparts and are implementable on QPGAs with significantly fewer layers. To better characterize this, we performed a search over qubit number and circuit depth to find trained circuits which match the target operator to within some specified fidelity threshold. We used the quantum Fourier transform as the target operator for this benchmark due to its prevalence and complexity. The results are plotted in Figure \ref{fig:circuit_depth}.

To perform the compactness analysis, we iteratively trained QPGAs of increasing depth to implement an $n$-qubit QFT to a desired fidelity threshold, chosen to be $\mathcal{F} > 99.9\%$. Multiple optimization routines were run at each depth since training does not converge deterministically due to random initialization and the potential for getting stuck in a local maxima, which is more pronounced at larger qubit numbers.\footnote{For circuits with many qubits, more sophisticated initialization routines which take the locally-connected structure of the architecture into account such as Haar initialization \cite{Pai2019MatrixDevices} may be necessary to ensure a reasonable chance of convergence.} We note that the final gradient-based QFT implementations typically require only $1/4$ to $1/3$ as many layers as their explicitly decomposed counterparts.

\section{Conclusion}
\label{sec:conclusion}

In this paper we have presented a photonic architecture for a quantum programmable gate array capable of implementing arbitrary quantum states, operators, and computations. The architecture, presented in Section \ref{sec:lattice_design}, extends universal programmable optics to the quantum domain by employing two-photon interactions from quantum emitters embedded in the waveguides. This allows for deterministic multi-qubit gates which use a number of waveguides that is linear in the number of qubits. The design parameterizes arbitrary quantum circuits as a lattice of single-qubit gates implemented by phase-modulated Mach-Zehnder interferometers and two-qubit $\cz$ gates with variable connectivity implemented by a scattering process described in Section \ref{sec:two_photon_gates}. By setting phase shifter parameters to implement appropriate single-qubit operations and to enable two-photon interactions where needed, the lattice can be dynamically programmed to implement any quantum circuit without hardware modifications.

In Section \ref{sec:implementing_quantum_operations}, we showed that the logical system implemented by the QPGA is computationally universal: any quantum operation can be mapped onto a corresponding set of phase shifter parameters given a sufficiently large lattice. We described an explicit algorithm for preparing arbitrary quantum states on the lattice which are efficient for some subclasses of quantum states, and we discussed how QR decomposition can map $\mathrm{U}(2^n)$ unitaries into a series of controlled rotations implementable on the lattice.

In Section \ref{sec:gradient-based-circuit-optimization}, we showed how gradient-based optimization techniques prevalent in machine learning can be used to automatically implement high-fidelity approximations to desired quantum operations. We trained simulated QPGAs with fixed $\cz$ connectivity to prepare a variety of important quantum states and operators, and we showed that these approximate circuit implementations are often significantly more compact than their explicitly-decomposed counterparts.

While this work is purely theoretical, there has been tremendous recent experimental progress in both of the key technologies required to realize this device: programmable photonic processors \cite{Shen2017DeepCircuits, Harris2018LinearProcessors, Harris2017QuantumProcessor, Han2015Large-scaleCouplers, Chung2017ACMOS, Ribeiro2016DemonstrationComponent, Zhuang2015ProgrammableApplications, Perez2017MultipurposeCore, Perez2018ProgrammableNanophotonics} and strongly coupled quantum emitters \cite{Zhang2018StronglyDiamond, Radulaski2019NanodiamondDevices, Laucht2012ASource, Kim2017HybridChip, Sollner2015DeterministicCircuits, Babinec2010ASource, Akimov2007GenerationDots, Bracher2017SelectiveCenter}. The ongoing advancements in these technologies may allow for feasible near-future implementation of the device described in this paper.

\section*{Acknowledgements}

The authors acknowledge helpful discussions with Sunil Pai, Casey Wojcik, Tyler Hughes, Avik Dutt, and Beicheng Lou. This work was supported by a Vannevar Bush Faculty Fellowship from the U.S. Department of Defense (Grant No. N00014-17-1-3030).

\bibliography{main.bbl}

% Appendices ====================================================================

\onecolumngrid
\appendix

\section{Phase-modulated interference for photons with arbitrary spectra}
\label{sec:interferometer_arbitrary_spectra}

\begin{figure}[b]
    \centering
    \includegraphics[width=0.5\textwidth]{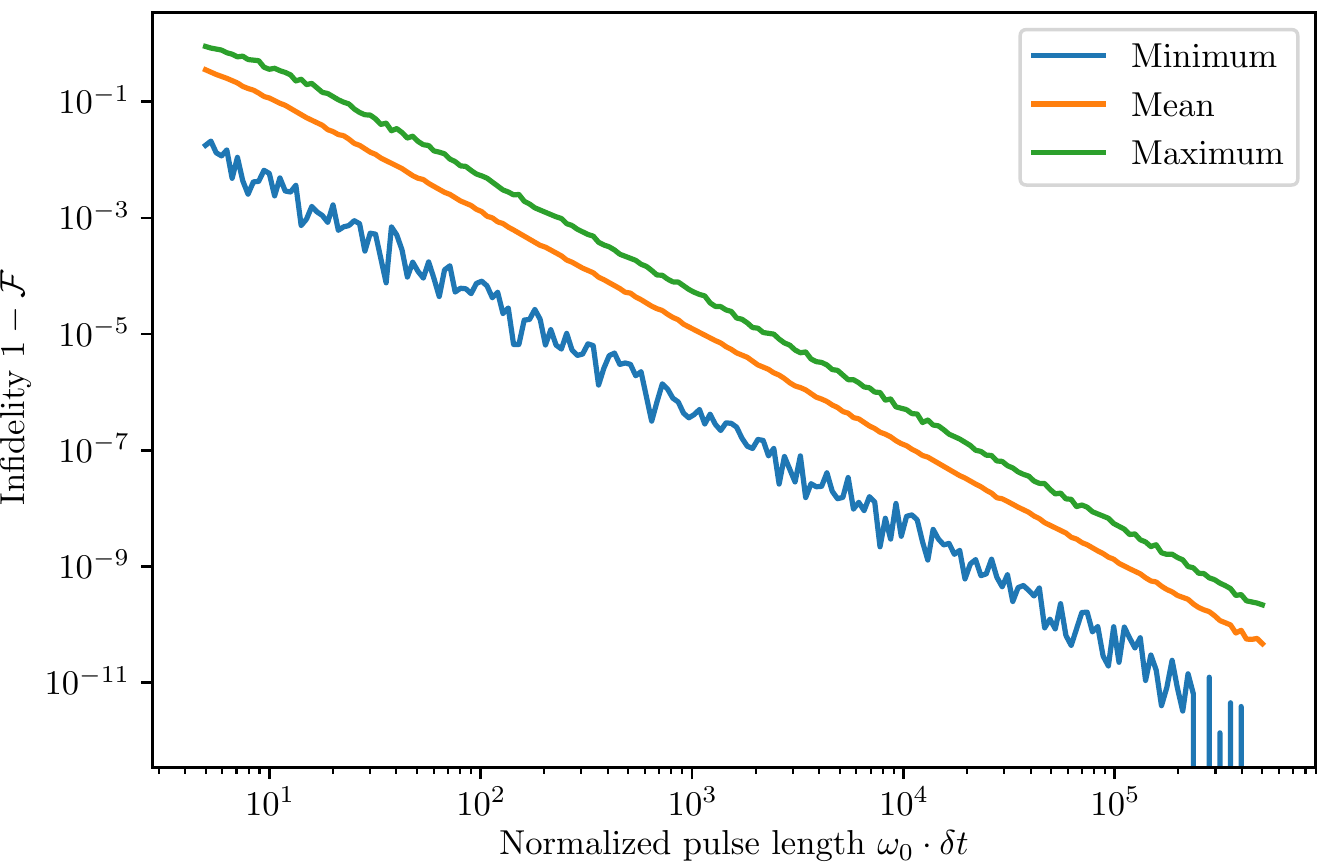}
    \caption{Infidelities of the output state of a Mach-Zehnder interferometer for a range of spectral distributions. We assume an input wavefunction of $\ket{\psi^\text{in}} = \int d \omega \, g_{\delta t} (\omega) \frac{1}{\sqrt{2}} (\adag_1 (\omega) + \adag_2 (\omega) ) \vac$, where $g_{\sigma} (\omega)$ is a Gaussian with a spectral width of $\sigma = \frac{\delta \omega}{\omega_0}$ and a pulse length of $\delta t = \frac{1}{2 \delta \omega}$ periods of the central frequency $\omega_0$. We compute the output wavefunction for an ensemble of 1000 values of $\zeta,\xi,\theta,\phi$ sampled uniformly from $[0,2\pi)$ across 250 values of $\sigma$ and plot the maximum, minimum, and average infidelity (defined as $1-\mathcal{F}$) for each case.}
    \label{fig:fidelity_mzi}
\end{figure}

Consider a Mach-Zehnder interferometer with four phase shifters in the arrangement presented in Figure \ref{fig:lattice}c. Let the operators $\adag_1 (\omega), \adag_2 (\omega)$ represent creation operators for the top and bottom waveguides, respectively, acting on a single frequency mode $\omega$. Consider an input state to the MZI representing a single logical qubit in the state $\alpha \ket{0_L} + \beta \ket{1_L}$:
\begin{equation}
\ket{\psi^\text{in}} = \int d \omega \, \phi (\omega) \left(\alpha \adag_1 (\omega) + \beta \adag_2 (\omega) \right) \vac.
\end{equation}

The phase shifters in the MZI act by imparting a time delay $\tau$ on the creation operators, mapping $\adag (\omega) \mapsto \adag(\omega) e^{i \omega \tau}$. (Here we make the approximation that the phase shifter imparts an equal time delay across the range of frequencies of the photon, e.g. has a constant refractive index.) Let $\{\tau_\zeta, \tau_\xi, \tau_\theta, \tau_\phi \} \equiv \{\zeta, \xi, \theta, \phi\} / \omega_0$ denote the effective time delays imparted by the four phase shifters, where $\omega_0$ denotes the 4LS resonance frequency $\omega$ in the main text. The idealized action of the MZI on photons of zero spectral width described in Eq. \ref{eq:mzi_four_phase_shifters} is $R^\alpha_\beta H R^\theta H R^\phi$. In the case of finite spectral width, the transformation maps:

\begin{equation}
\label{eq:mzi_operator_mapping}
\begin{bmatrix}
\adag_1 (\omega) \\
\adag_2 (\omega)
\end{bmatrix}
\leftarrow
\frac{1}{2} 
\begin{bmatrix}
e^{i (\tau_\zeta +\tau_\phi ) \omega} \left(e^{i \tau_\theta \omega} + 1\right) & e^{i (\tau_\xi + \tau_\phi ) \omega} \left(e^{i \tau_\theta \omega} - 1\right) \\
e^{i \tau_\zeta \omega} \left(e^{i \tau_\theta \omega} - 1\right) & e^{i \tau_\xi \omega} \left(e^{i \tau_\theta \omega} + 1\right) \\
\end{bmatrix}
\begin{bmatrix}
\adag_1 (\omega) \\
\adag_2 (\omega)
\end{bmatrix}.
\end{equation}

Thus, the output state of the MZI is:
\begin{multline}
\ket{\psi^\text{out}} = \frac{1}{2} \int d \omega \, \phi (\omega) 
\left[
\left(
\alpha e^{i (\tau_\zeta +\tau_\phi ) \omega} \left(e^{i \tau_\theta \omega} + 1\right) + 
\beta e^{i (\tau_\xi + \tau_\phi ) \omega} \left(e^{i \tau_\theta \omega} - 1\right)
\right) \adag_1 (\omega) 
\right. \\
+ 
\left.
\left( 
\alpha e^{i \tau_\zeta \omega} \left(e^{i \tau_\theta \omega} - 1\right) +
\beta e^{i \tau_\xi \omega} \left(e^{i \tau_\theta \omega} + 1\right)
\right) \adag_2 (\omega) 
\right]
\vac.
\end{multline}

Define coefficients $C_0 (\omega) \equiv \frac{1}{2} ( \alpha e^{i (\tau_\zeta +\tau_\phi ) \omega} \left(e^{i \tau_\theta \omega} + 1\right) + \beta e^{i (\tau_\xi + \tau_\phi ) \omega} \left(e^{i \tau_\theta \omega} - 1\right) )$ and $C_1 (\omega) \equiv \frac{1}{2} (\alpha e^{i \tau_\zeta \omega} \left(e^{i \tau_\theta \omega} - 1\right) + \beta e^{i \tau_\xi \omega} \left(e^{i \tau_\theta \omega} + 1\right) )$. Then the output state is $\ket{\psi^\text{out}} = \int d \omega \, \phi(\omega) (C_0 (\omega) \adag_1 (\omega) + C_1 (\omega) \adag_2 (\omega)) \vac$. Define projection operators $\hat{P}_0, \hat{P}_1$ which map physical wavefunctions to logical state vectors:

\begin{align}
\hat{P}_0 &= \int d\omega \ket{0_L} \bra{\emptyset} \hat{a}_1 (\omega) \\
\hat{P}_1 &= \int d\omega \ket{1_L} \bra{\emptyset} \hat{a}_2 (\omega) 
\end{align}

To obtain the fidelity of the physical output state against the target logical output state $\ket{\psi^\text{targ}} = C_0^L \ket{0} + C_1^L \ket{1}$, we evaluate the inner product between the states by expanding in terms of the complete basis $\id = \hat{P}_0 + \hat{P}_1$: 

\begin{align}
\begin{split}
\mathcal{F} &= \abs{\braket{\psi^\text{targ} | \psi^\text{out}}}^2 \\
&= \left| \left(C_0^{L*} \bra{0_L} + C_1^{L*} \bra{1_L}\right) \int d \omega \, \phi(\omega) \left( C_0 (\omega) \adag_1 (\omega) + C_1 (\omega) \adag_2 (\omega) \right) \vac \right|^2 \\
&= \left| \left(C_0^{L*} \bra{0_L} + C_1^{L*} \bra{1_L}\right) \left(\hat{P}_0 + \hat{P}_1\right) \int d \omega \, \phi(\omega) \left( C_0 (\omega) \adag_1 (\omega) + C_1 (\omega) \adag_2 (\omega) \right) \vac \right|^2 \\
&= \left| \int d\omega \, \phi(\omega) \left( C_0^{L*} C_0 (\omega) + C_1^{L*} C_1 (\omega) \right)\right|^2.
\end{split}
\end{align}

The fidelity of the output state will depend on the phase shifter values. We numerically simulate the output wavefunctions for a large sample of $\zeta,\xi,\theta,\phi$ across a range of spectral widths and plot the results in Figure \ref{fig:fidelity_mzi}.

\section{Derivation of reflection coefficients}
\label{sec:derivation_of_reflection_coefficients}

In this section, we derive the reflection coefficients presented in Section \ref{sec:two_photon_gates}, using a similar treatment of the problem as in Ref. \cite{Zheng2013Waveguide-QED-basedComputation}. To simplify the derivation, we replace the Hamiltonian in Eq. \ref{eq:hamiltonian} with an \emph{ad-hoc} Hamiltonian:
\begin{multline}
\label{eq:adhoc_hamiltonian}
\mathcal{H}_\text{ad-hoc} = \frac{\hbar}{i} \int dx \left[ \hat{b}_R^\dag (x) \frac{\partial}{\partial x} \hat{b}_R (x) - \hat{b}_L^\dag (x) \frac{\partial}{\partial x} \hat{b}_L (x) \right] + \hbar \sum_{n=2}^{4} \left( \Omega_n - \frac{i \Gamma'}{2} \right) \ketbra{n}{n} \\
+ \hbar \int dx \, \sqrt{\Gamma/2} \, \delta(x) \left[ \left(\hat{b}_R^\dag (x) + \hat{b}_L^\dag (x) \right) \left(\ketbra{1}{2} + \ketbra{3}{2} + \ketbra{3}{4} \right) + \text{H.c.} \right],
\end{multline}
where we have also set $v_g = v_r = 1$. With this approach, the Hilbert space contains only waveguide and atom states, without the environmental reservoir. This \emph{ad-hoc} approach is known to produce correct scattering matrices for single-photon (and temporally-separated multi-photon) interactions, and is thus suitable for our purposes, but it should be noted that the direct substitution of $\Omega \rightarrow \Omega - i \Gamma' / 2$ in the Hamiltonian rather than in the scattering matrix will yield incorrect results for temporally overlapping two-photon scattering. \cite{Rephaeli2013DissipationInvited}

\emph{Step 1.} 
Consider the dynamics of a single quantum emitter in the device, from sites (4a) to (6a) in Fig. \ref{fig:lattice}c. Photon $A$ at frequency $\omega=\Omega_{12} = \Omega_{34}$ is incident on the atom, which is initialized in state $\ket{1}$. The stationary state of the system is:
\begin{equation}
    \ket{\psi_1} = \int dx \left[ \phi_{1R}(x) \bdag_R (x) + \phi_{1L}(x) \bdag_L (x) \right] \vac \otimes \ket{1} + e_2 \vac \otimes \ket{2} + \int dx \left[ \phi_{3R}(x) \bdag_R (x) + \phi_{3L}(x) \bdag_L (x) \right] \vac \otimes \ket{3},
\end{equation}
where the amplitude of the $\phi$ wavepackets correspond to the component of the photon which is in the spatial mode being considered. \cite{Shen2005CoherentWaveguides, Zheng2013Waveguide-QED-basedComputation} Using the Schrodinger equation $ \mathcal{H} \ket{\psi_1} = \hbar \omega \ket{\psi_1}$, where $\mathcal{H}$ is given in Eq. \ref{eq:adhoc_hamiltonian}, and defining a coupling constant $V \equiv \sqrt{v_g \Gamma / 2}$ we obtain:
\begin{subequations}
\label{eq:equations_of_motion_photon1}
\begin{align}
\left( -i \frac{d}{dx} - \omega \right) \phi_{1R} (x) + V \delta(x) e_2 &= 0, \\
\left( +i \frac{d}{dx} - \omega \right) \phi_{1L} (x) + V \delta(x) e_2 &= 0, \\
\left( -i \frac{d}{dx} - \omega' \right) \phi_{3R} (x) + V \delta(x) e_2 &= 0, \\
\left( +i \frac{d}{dx} - \omega' \right) \phi_{3L} (x) + V \delta(x) e_2 &= 0, \\
-\frac{i \Gamma'}{2} e_2 + V \left( \phi_{1R}(0) + \phi_{1L}(0) + \phi_{3R}(0) + \phi_{3L}(0) \right) &= 0.
\end{align}
\end{subequations}
Defining $k \equiv \omega / c$ and $k' \equiv \omega' / c = \Omega_{32} / c$, and following the treatment in Ref. \cite{Shen2005CoherentWaveguides} and \cite{Zheng2013Waveguide-QED-basedComputation}, we assume a solution ansatz of:
\begin{subequations}
\label{eq:ansatz_photon1}
\begin{align}
\phi_{1R} (x) &= e^{+i k x} \left( \theta(-x) + \beta_{1R} \theta(x) \right), \\
\phi_{1L} (x) &= e^{-i k x} \left(\alpha_{1L} \theta(-x) + \beta_{1L} \theta(x) \right), \\
\phi_{3L} (x) &= e^{-i k' x} \left(\beta_{3L} \theta(-x) \right), \\
\phi_{3R} (x) &= e^{+i k' x} \left(\beta_{3L} \theta(-x) + \alpha_{3R} \theta(x) \right),
\end{align}
\end{subequations}
where $\theta$ is the Heaviside function with $\theta(0)\equiv \frac{1}{2}$. Here, $\beta$ coefficients describe parts of the wavefunction between the relevant reflector and the 4LS, while $\alpha$ coefficients describe parts which are outside the 4LS (the input/output waveguide for the $\omega$ photon and the delay line for the $\omega'$ photon). The reversal of direction of $x$ for $\phi_1$ and $\phi_3$ is due to the opposite orientation of the reflectors for $\omega$ and $\omega'$, respectively. The reflective boundary conditions at $x=\pm a$ means that:
\begin{equation}
    \phi_{1R}(a)+\phi_{1L}(a) = 0 = \phi_{3L}(-a) + \phi_{3R}(-a).
\end{equation}
Using this and substituting equations \ref{eq:ansatz_photon1} into \ref{eq:equations_of_motion_photon1} gives us the solution:
\begin{align}
r_{11} = \alpha_{1L} &= e^{2i \omega a} \frac
{\frac{i\Gamma'}{2} - \frac{i\Gamma}{2} \left( e^{2i \omega' a} - e^{-2i \omega a} \right)}
{-\frac{i\Gamma'}{2} + \frac{i\Gamma}{2} \left( e^{2i \omega' a} + e^{2i \omega a} - 2 \right)}, \label{eq:r11}\\
r_{13} = \alpha_{3R} &= \frac
{\frac{i\Gamma}{2} \left( e^{2i\omega a} - 1 \right) \left( e^{2i\omega' a} - 1 \right)}
{-\frac{i\Gamma'}{2} + \frac{i\Gamma}{2} \left( e^{2i \omega' a} + e^{2i \omega a} - 2 \right)}. \label{eq:r13}
\end{align}

\emph{Step 2.} 
We now send in the second photon $B$, also of frequency $\omega$, which will scatter off of the $\ket{1}$ component of the 4LS state in the same manner as the first photon. We assume that the temporal separation of photons $A$ and $B$ is much greater than the decay timescale of the excited $\ket{2}, \ket{4}$ states, and since $\omega$ is off resonance from the $\ket{3} \leftrightarrow \ket{2}$ transition at $\omega'$, then $B$ will interact with the $\ket{3} \leftrightarrow \ket{4}$ transition only. The single photon scattering eigenstate for the $\ket{3}$ component of the 4LS state then takes the form:
\begin{equation}
    \ket{\psi_2} = \int dx \left[ \phi_{3R}(x) \bdag_R (x) + \phi_{3L} (x) \bdag_L (x) \right] \vac \otimes \ket{3} + e_4 \vac \otimes \ket{4}.
\end{equation}
As before, applying the \emph{ad-hoc} Hamiltonian to $ \mathcal{H} \ket{\psi_2} = \hbar \omega \ket{\psi_2}$, we obtain equations of motion:
\begin{subequations}
\label{eq:equations_of_motion_photon2}
\begin{align}
\left( -i \frac{d}{dx} - \omega \right) \phi_{3R} (x) + V \delta(x) e_4 &= 0, \\
\left( +i \frac{d}{dx} - \omega \right) \phi_{3L} (x) + V \delta(x) e_4 &= 0, \\
-\frac{i \Gamma'}{2} e_4 + V \left( \phi_{3R}(0) + \phi_{3L}(0) \right) &= 0.
\end{align}
\end{subequations}
Assuming a solution ansatz of
\begin{subequations}
\label{eq:ansatz_photon2}
\begin{align}
\phi_{3R} (x) &= e^{+i k x} \left( \theta(-x) + \beta_{3R} \theta(x) \right), \\
\phi_{3L} (x) &= e^{-i k x} \left(\alpha_{3L} \theta(-x) + \beta_{3L} \theta(x) \right), 
\end{align}
\end{subequations}
where $k$ is defined as before, and imposing reflective boundary conditions that $\phi_{3R}(a) + \phi_{3L}(a) = 0$, we obtain the reflected amplitude to be:
\begin{equation}
R_3 = \alpha_{3L} = \frac
{\frac{i\Gamma'}{2}e^{2i\omega a} + \frac{i\Gamma}{2} \left(1-e^{2i\omega a}\right)}
{-\frac{i\Gamma'}{2} - \frac{i\Gamma}{2} \left(1-e^{2i\omega a}\right)}.
\end{equation}

\emph{Step 3.} 
The $A'$ photon of frequency $\omega'$ has traveled down the delay line and back and is incident on the 4LS, which is in some superposition of $\ket{1}$ and $\ket{3}$. The photon is far off-resonance from the $\ket{1}\leftrightarrow\ket{2}$ transition, so will only interact with the $\ket{3}\leftrightarrow\ket{2}$ transition. Using the same approach as before, we obtain reflection amplitudes which are analogous to Eqs. \ref{eq:r11} and \ref{eq:r13}, except with $\omega$ and $\omega'$ switched:
\begin{align}
r_{33} &= e^{2i \omega' a} \frac
{\frac{i\Gamma'}{2} - \frac{i\Gamma}{2} \left( e^{2i \omega a} - e^{-2i \omega' a} \right)}
{-\frac{i\Gamma'}{2} + \frac{i\Gamma}{2} \left( e^{2i \omega a} + e^{2i \omega' a} - 2 \right)}, \label{eq:r33}\\
r_{31} &= \frac
{\frac{i\Gamma}{2} \left( e^{2i\omega' a} - 1 \right) \left( e^{2i\omega a} - 1 \right)}
{-\frac{i\Gamma'}{2} + \frac{i\Gamma}{2} \left( e^{2i \omega a} + e^{2i \omega' a} - 2 \right)}. \label{eq:r31}
\end{align}

\emph{Step 4.} 
The $B'$ photon of frequency $\omega'$ has returned to the 4LS, which is in some different superposition of $\ket{1}$ and $\ket{3}$. As before, the photon only interacts with the $\ket{3}\leftrightarrow\ket{2}$ transition, and has identical reflection coefficients as step 3.

\newpage 

\section{Implementations of common quantum gates}
\label{sec:gate_implementations}

\begin{table*}[ht]
\setlength{\tabcolsep}{.5em} % for the horizontal padding
\begin{tabular}{llcccc}
\toprule
Operator & Matrix representation & $\zeta$ & $\xi$ & $\theta$ & $\phi$
\\\midrule
Identity & $\id = \smat{1}{0}{0}{1}$ & 
$0$ & $0$ & $0$ & $0$ \\
Hadamard & $H=\frac{1}{\sqrt{2}}\smat{1}{1}{1}{-1}$ & 
$\frac{5\pi}{4}$ & $\frac{3\pi}{4}$ & $\frac{\pi}{2}$ & $\frac{\pi}{2}$ \\
Pauli-X & $\sigma_x = \smat{0}{1}{1}{0}$ & 
$\pi$ & $\pi$ & $\pi$ & $0$ \\
Pauli-Y & $\sigma_y = \smat{0}{-i}{i}{0}$ & 
$\frac{3\pi}{2}$ & $\frac{\pi}{2}$ & $\pi$ & $0$ \\
Pauli-Z & $\sigma_z = \smat{1}{0}{0}{-1}$ & 
$0$ & $\pi$ & $0$ & $0$ \\
Rotation-X & $R_x(\theta') = \cos\frac{\theta'}{2}\id - i \sin\frac{\theta'}{2}\sigma_x$ & 
$-2\theta'$ & $-2\theta'$ & $\theta'$ & $0$ \\
Rotation-Y & $R_y(\theta') = \cos\frac{\theta'}{2}\id - i \sin\frac{\theta'}{2}\sigma_x$ & 
$-2\theta'-\frac{\pi}{2}$ & $-2\theta'$ & $\theta'$ & $\frac{\pi}{2}$ \\
Rotation-Z & $R_z(\theta') = \cos\frac{\theta'}{2}\id - i \sin\frac{\theta'}{2}\sigma_z$ & 
$-\frac{\theta'}{2}$ & $\frac{\theta'}{2}$ & $0$ & $0$ \\
Phase shift & $R_{\phi'} = \smat{1}{0}{0}{e^{i\phi}}$ & 
$0$ & $\phi'$ & $0$ & $0$ \\
\bottomrule
\end{tabular}
\caption{A table of phase shifter parameters which implement various common single-qubit gates on the phase-modulated MZIs depicted in Figure \ref{fig:lattice}c.}
\label{table:single_qubit_gates}
\end{table*}

\begin{table*}[ht]
\setlength{\tabcolsep}{.5em} % for the horizontal padding
\begin{tabular}{lccc}
\toprule
Operator & Symbol & Decomposition & Required $\cz$ layers
\\\midrule
Identity & 
\tab{\Qcircuit@C=1em@R=1.2em{&\qw&\qw&\qw\\&\qw&\qw&\qw}} & 
\tab{\Qcircuit@C=1em@R=.7em{&\gate{I}&\ctrl{1}&\gate{I}&\ctrl{1}&\gate{I}&\qw\\&\gate{I}&\ctrl{-1}&\gate{I}&\ctrl{-1}&\gate{I}&\qw}} & 
$0, 2, 4, \cdots$ \\[2em]
Controlled-NOT & 
\tab{\Qcircuit@C=1em@R=1.2em{&\ctrl{1}&\qw\\&\targ&\qw}} & 
\tab{\Qcircuit@C=1em@R=.7em{&\gate{I}&\ctrl{1}&\gate{I}&\qw\\&\gate{H}&\ctrl{-1}&\gate{H}&\qw}} & 
1 \\[2em]
Controlled-phase & 
\tab{\Qcircuit@C=1em@R=1.2em{&\ctrl{1}&\qw\\&\gate{R_\phi}&\qw}} & 
\tab{\Qcircuit@C=1em@R=.7em{&\gate{R_{\phi/2}}&\ctrl{1}&\gate{I}&\ctrl{1}&\gate{I}&\qw\\&\gate{R_z(\phi)H}&\ctrl{-1}&\gate{H R_z(-\frac{\phi}{2}) H}&\ctrl{-1}&\gate{H R_z(\frac{\phi}{2})}&\qw}} & 
2 \\[2em]
Controlled-$U$\footnote{We decompose $U$ via Euler angles as $U=e^{i\phi} R_z(\alpha)R_y(\theta)R_z(\beta)$.} &
\tab{\Qcircuit@C=1em@R=1.2em{&\ctrl{1}&\qw\\&\gate{U}&\qw}} & 
\tab{\Qcircuit@C=1em@R=.7em{&\gate{R_{\phi}}&\ctrl{1}&\gate{I}&\ctrl{1}&\gate{I}&\qw\\&\gate{R_z(\alpha)R_y(\frac{\theta}{2})H}&\ctrl{-1}&\gate{HR_y(-\frac{\theta}{2})R_z(-\frac{\alpha + \beta}{2})H}&\ctrl{-1}&\gate{HR_z(\frac{\beta-\alpha}{2})}&\qw}} & 
2 \\[2em]
SWAP & 
\tab{\Qcircuit@C=1em@R=1.2em{&\qswap&\qw\\&\qswap\qwx&\qw}} & 
\tab{\Qcircuit@C=1em@R=.7em{&\gate{H}&\ctrl{1}&\gate{H}&\ctrl{1}&\gate{H}&\ctrl{1}&\qw\\&\gate{H}&\ctrl{-1}&\gate{H}&\ctrl{-1}&\gate{H}&\ctrl{-1}&\qw}} & 
3 \\[2em]
\bottomrule
\end{tabular}
\caption{Construction of common multi-qubit gates by embedding single-qubit operations in a lattice of $\cz$ gates. Because the phase-modulated MZIs can implement any single-qubit operator, gate decompositions may be terminated with either with $\cz$ gates or with single-qubit gates, as the first layer of single-qubit operators of subsequent gates can implicitly include the final single-qubit operators of the previous logical gate. All quantum circuit diagrams in this paper were typeset using the \texttt{QCircuit} \LaTeX package. \cite{Eastin2004Q-circuitTutorial}}
\label{table:gates}
\end{table*}

\end{document}